\documentclass[aps,prl,twocolumn,preprintnumbers,superscriptaddress,amsmath]{revtex4}
\usepackage{graphicx}

\usepackage{amsmath}
\usepackage{amssymb}
\usepackage{enumitem}
\usepackage{multirow}
\usepackage{physics}
\usepackage[dvipsnames]{xcolor}
\usepackage[T1]{fontenc}
\usepackage{graphicx}
\usepackage{dcolumn}
\usepackage{bm}
\usepackage[retainorgcmds]{IEEEtrantools}
\usepackage{xcolor}
\usepackage[colorlinks=true,linkcolor=blue,citecolor=blue,urlcolor=blue]{hyperref}
\usepackage{ulem}
\usepackage{bbold}
\usepackage{mathtools}
\usepackage{accents}
\usepackage{verbatim}
\usepackage{gensymb}
\usepackage[commandnameprefix=always]{changes}

\usepackage{fp}
\newcount\boxheight
\newcount\boxwidth
\newcommand\testaspect[1]{%
  \setbox0=\hbox{#1}%
  \boxheight=\ht0\relax%
  \boxwidth=\wd0\relax%
  \FPdiv\theaspect{\the\boxwidth}{\the\boxheight}%
  \copy0%
}
\makeatletter
\@namedef{Changes@AuthorColor}{magenta}
\colorlet{Changes@Color}{magenta}
\makeatother

\definecolor{light_blue}{HTML}{17BECF}
\definecolor{blue}{HTML}{1F77B4}
\definecolor{green}{HTML}{2CA02C}
\definecolor{yellow}{HTML}{FAC205}
\definecolor{orange}{HTML}{FF7F0E}
\definecolor{pink}{HTML}{E377C2}
\definecolor{brown}{HTML}{8C564B}
\definecolor{light_red}{HTML}{EF4026}
\definecolor{violet}{HTML}{7E1E9C}

\newcommand{\xc}{\mathrm{xc}}

\newcommand{\Hxc}{\mathrm{Hxc}}

\newcommand{\ext}{\mathrm{ext}}

\newcommand{\p}{^\prime}

\newcommand{\indxlt}{\Gamma/\mathrm{K}}
\newcommand{\GDFT}{\mathrm{GDFT}}

\DeclareMathAlphabet{\mathsfit}{\encodingdefault}{\sfdefault}{m}{sl}
\SetMathAlphabet{\mathsfit}{bold}{\encodingdefault}{\sfdefault}{bx}{sl}

\newcommand{\w}{\omega}

\newcommand{\F}{\mathrm{F}}

\newcommand{\DFT}{\mathrm{DFT}}

\makeatletter
\def\maketitle{
\@author@finish
\title@column\titleblock@produce
\suppressfloats[t]}
\makeatother

\begin{document}
\title{Excitonic effects in phonons: reshaping the graphene Kohn anomalies and lifetimes}

\author{Alberto Guandalini}
\email{alberto.guandalini@uniroma1.it}
\affiliation{Dipartimento di Fisica, Universit\`a di Roma La Sapienza, Piazzale Aldo Moro 5, I-00185 Roma, Italy}

\author{Francesco Macheda}
\affiliation{Dipartimento di Fisica, Universit\`a di Roma La Sapienza, Piazzale Aldo Moro 5, I-00185 Roma, Italy}

\author{Giovanni Caldarelli}
\affiliation{Dipartimento di Fisica, Universit\`a di Roma La Sapienza, Piazzale Aldo Moro 5, I-00185 Roma, Italy}

\author{Francesco Mauri}
\affiliation{Dipartimento di Fisica, Universit\`a di Roma La Sapienza, Piazzale Aldo Moro 5, I-00185 Roma, Italy}

\begin{abstract}
We develop an ab initio framework that captures the impact of electron-electron and electron-hole interactions on phonon properties.
  This enables the inclusion of excitonic effects in the optical phonon dispersions and lifetimes of graphene, both near the center ($\Gamma$) and at the border (K) of the Brillouin zone, at phonon-momenta relevant for Raman scattering and for the onset of the intrinsic electrical resistivity. Near K, we find a phonon red-shift of $\sim 150$ cm$^{-1}$ and a 10x enhancement of the group velocity, together with a $5 \times$ increase in linewidths due to a $26 \times$ increase of the electron-phonon matrix elements. These effects persist for doping $2E_{\F} < \hbar \omega_{\mathrm{ph}}$ and are quenched at higher dopings.  Near $\Gamma$, the excitonic effects are minor because of the gauge field nature of the electron-phonon coupling at small phonon momentum. 
\end{abstract}

\maketitle
Experimental phonon dispersions of graphite~\cite{Jishi_1982,Aizawa_1990,Maultzsch_2004,Mohr_2007,Saito_2001} are generically in good agreement with density functional theory (DFT) predictions, with the remarkable exception of measured  transverse optical branch near the K point~\cite{Gruneis_2009}, which reveals a redshift of the phonon frequencies and an increase of the group velocity with respect to the DFT results. A satisfactory comparison with experiments is restored via the explicit inclusion of electron-electron interaction effects~\cite{Lazzeri_2008,Gruneis_2009,Attaccalite_2009}, not accounted in DFT, via a frozen phonon approach combined with the GW approximation~\cite{Hedin_1965,Strinati_1982,Hybertsen_1986,Godby_1988}.

The interaction effects are mediated by the screened Coulomb $W$ which is larger in a monolayer than in a thick multilayer. Therefore, relevant differences between graphene and graphite phonons are expected, especially at the K point as recently hinted via infrared Raman measurement~\cite{Venanzi_2023,Graziotto2024-zk,PhysRevB.109.075420}.

In this work, we study dispersions and linewidths of transverse and longitudinal optical phonons (TO, LO) of graphene, in the regions near the $\Gamma$ and K points, relevant for the electron-phonon scattering ruling Raman spectra and the electrical resistivity. 
We include electronic interaction effects in the dynamical matrix in the framework of generalized {Kohn-Sham} theory (GDFT)\cite{bauernschmitt1996treatment,rocca2008turbo,Baer2018GenTDDFT}, apt to reproduce excitonic effects at same level of the Bethe-Salpeter equation (BSE)~\cite{Hedin_65,Strinati_88,Onida_02} with static $W$. We evaluate the effects of the electron-hole interaction on the response as an infinite summation of electronic ladder diagrams (excitonic effects) using a tight-binding model which accurately reproduces the low-energy physics of graphene \cite{albertoTB}. 
We further leverage on the recent variational formulation of the phonon response in the presence of nonlocal exchange interactions~\cite{Giovanni_2024} to maximize the accuracy of the approximations. 
 
Previous efforts evaluated the impact of interaction effects on the electron-phonon matrix elements by considering the variation of the one-particle electronic Green's function{, even with GW~\cite{Attaccalite_2009,Li_2019} or the renormalization-group approach~\cite{Basko_2008,Basko_2009},} then computing the phonon self-energy in an effective non-interacting framework with interacting vertices. %{or using a renormalization group approach [Ref. Basko]}. 
We instead obtain the phonon self-energy from the interacting two-particle electron Green's function, including interactions beyond finite-order perturbation theory. With the proposed approach, we provide a formally exact theoretical framework where the sole approximation pertains to the treatment of the electronic correlation.
Exploiting the exact generalized Fermi golden rule derived in Ref.\ \cite{Giovanni_2024}, we finally extract effective electron-phonon matrix elements including excitonic effects, which are object of recent research ~\cite{laricchia2024}, but their impact on the phonon spectral function has not been addressed.

We find a strong redshift of the freestanding graphene phonon dispersion at the K point with respect to graphite, with an associated enhancement of the group velocity, as well as a strong increase of the linewidths. Treating the full infinite summation of electronic ladder diagrams is crucial to obtain quantitative results. We also investigate dynamical and doping effects, which are known to be relevant for optical graphene phonons near $\Gamma$ and K~\cite{Lazzeri_2006,Popov_2010}, and evaluate their impact on the magnitude of the enhancement.

\paragraph{Phonon Green's function---}
We consider a spin unpolarized finite system of $N_{\text{ion}}$ ions and $N_{\text{el}}$ electrons interacting through the Coulomb potential. The generalization to periodic systems, e.g.\ graphene, is straightforward, and covered in the Supplemental Material (SM)~\cite{supp-info}. We use the convention that integration over the volume of the system for spatial variables with numbers as a subscript is implicit, as in Ref.~\onlinecite{Giovanni_2024}.
We express the analytic continuation of Fourier transforms of retarded quantities in complex frequency $z{=}\omega {+} i \eta$. Physical observables are obtained in the limit $\eta{\to}0^+$, to be considered after the thermodynamical limit~\cite{kubo2012statistical}.
All the information about phonons is encoded in the (mass-reduced) displacement-displacement Green's function~\cite{maximov75,allen79,allen-book,Giustino_2017}
\begin{equation}\label{phonon green function}
     G^{-1}{\!\!\!\,}_{s\alpha,r\beta}(\omega) = \omega^2\delta_{s\alpha,r\beta} {-} D_{s\alpha,r\beta}(\omega+i0^+) ,
\end{equation}
where $D$ is the dynamical matrix. Latin and Greek symbols indicate atomic and Cartesian coordinates respectively.
The exact evaluation of the phonon Green's function, in the harmonic approximation, requires the solution of the Hedin-Baym equations~\cite{Giustino_2017,Stefanucci_2023}.
We disregard electron-phonon effects on the electronic dynamics, such as the phonon vertex corrections neglected leveraging the Migdal theorem~\cite{migdal58} and electron-phonon insertions in the electronic Green's function.
Then, we obtain the phonon Green's function from the interacting electronic density induced by the (time-dependent) lattice perturbation at the equilibrium positions~\cite{Giovanni_2024}, or, equivalently, from the electron density-density response function.
Within this framework, the dynamical matrix can be expressed as
\begin{equation}\label{eq:D_def}
    D_{s\alpha,r\beta}(z) =\frac{C_{s\alpha,r\beta}(z)+C^{\mathrm{ion-ion}}_{s\alpha,r\beta}-\sum_{l} C_{s\alpha,l\beta}(z)\delta_{rs} }{\sqrt{M_sM_r}}
\end{equation}
where $M_s$/$M_r$ are the ionic masses of the $s$/$r$ atom. $C^{\mathrm{ion-ion}}_{s\alpha,r\beta}$ is the second derivative of the ion-ion interaction energy and
\begin{equation}\label{eq:C_def}
    C_{s\alpha,r\beta}(z) = V_{\ext}^{(s\alpha)}(\bm r_1)
    \chi(\bm r_1,\bm r_2,z)
    V_{\ext}^{(r\beta)}(\bm r_2),
\end{equation}
 where $V_{\ext}^{(s\alpha)}$ is the electron-ion and ion-ion interaction potentials derived with respect to the ion displacement \cite{Calandra_10}. $\chi$ is the electronic density-density response function with nuclei clamped at their equilibrium positions, defined by $\rho^{(s\alpha)}(\bm r, z) {=} \chi(\bm r, \bm r_1,z) V_{\ext}^{(s\alpha)}(\bm r_1)$, where $\rho^{(s\alpha)}$ is the  electronic density induced by the ion displacement.
 A derivation of Eq.~\eqref{eq:D_def} is shown in the SM~\cite{supp-info}.

\paragraph{Density response with excitonic effects---}The evaluation of Eq.\ \eqref{eq:D_def} requires the  density-density response function $\chi$, obtained e.g. with Hedin's equations~\cite{Hedin_65} and linear response theory.
However, practical schemes require approximations for the electronic interaction.
GW+BSE approaches accurately describe excitonic effects of low-energy excitations~\cite{Reining_16}, thus providing an accurate imaginary part of $\chi(z)$ in a wide range of crystals, including graphene~\cite{Yang2009,guandalini_2023,Guandalini_2024}.
This is done by employing a dynamically-screened interaction in the GW band structure and a static screened interaction $W(\bm r,\bm r') {=} \epsilon^{-1}(\bm r,\bm r_1)v(|\bm r_1{-}\bm r'|)$ in the BSE, with $\epsilon^{-1}$ being the static inverse dielectric function in the random-phase approximation and $v(|\bm r|) {=} e^2/|\bm r|$ the Coulomb interaction.
However, their applicability to phonons, where also the real part of $\chi(z)$ is required in Eq.~\eqref{eq:C_def}, is still a computational challenge~\cite{Sondersted_24} with unknown accuracy.

GDFT with range-separated (RS) hybrid functionals ~\cite{Yoshinobu2008RangeSep1,Seth2012RangeSep2,Zhan_23} {are} extensively used to describe molecular vibrations with nonlocal exchange effects in linear response~\cite{jimenez2008evaluation}.
Within this framework, electrons and holes effectively interact with an empirical $W^{\mathrm{RS}}(\bm r{-}\bm r') {=} \alpha(\bm r-\bm r')v(\bm r {-} \bm r')$, where $\alpha$ is a combination of error functions weighting fractions of short and long-range exchange.
Dielectric-dependent hybrid functionals~\cite{Pasquarello_2023} aim to mimic the long-range $W$ of the BSE with $\alpha(|\bm r-\bm r'| \gg 1) = 1/\epsilon_{\infty}$, where $\epsilon_{\infty}$ is the electronic static dielectric constant.

In this work, we consider the following Hartree-exchange-correlation (Hxc) functional
\begin{align}
    E^{\GDFT}_{\Hxc}[n] {=} E^{\DFT}_{\Hxc}[\rho]{-}
    \frac{1}{2}|n(\bm r_1,\bm r_2)|^2\mathcal{W}(\bm r_1,\bm r_2){+}\bar{E}_{\xc}[\rho,\mathcal{W}] \label{eq:EHxc_def}\ ,
\end{align}
where $n(\bm r,\bm r\p) {=} \sum_i f_i \psi_i(\bm r)\psi^*_i(\bm r\p)$ is the (one-body) electronic density matrix, while $f_i$ and $\psi_i$ are Kohn-Sham occupations and orbitals. We use a screened interaction $\mathcal{W}(\bm r,\bm r')$ such that $\mathcal{W}(\bm r, \bm r') {=} W(\bm r,\bm r')$ for $|\bm r-\bm r'| {>} r_c$, with $r_c$ a cutoff radius in the range of interatomic distance.
For $|\bm r {-}\bm r'| {<} r_c$, $\mathcal{W}$ is smoothly reduced as $|\bm r {-}\bm r'| {\to} 0$, as in Refs.~\onlinecite{Gygi_1989,Massidda_1995,Rohlfing_2010}, in order to minimize the impact of the static screening approximation on the band structure.
$E^{\DFT}_{\Hxc}[\rho]$ is the $\Hxc$ energy in a local/semilocal approximation, e.g.\ the local density (LDA~\citep{LDA}) or generalized gradient approximation (GGA\cite{Perdew_1996}).
$\bar{E}_{\xc}[\rho,\mathcal{W}]$ {must be added to cancel the Fock-like term in Eq.~\eqref{eq:EHxc_def}, so}
%is a double counting term  (due to the presence of the Fock-like term , see SM~\cite{supp-info}), which is removed 
to recover the exact interacting total energy in the uniform  density limit (jellium). By setting $\mathcal{W}{=}0$ in Eq.~\eqref{eq:EHxc_def}, we recover the DFT functionals (LDA or GGA). 
Via time-dependent density functional theory~\cite{Baroni_2014} applied to the $E_{\Hxc}^{\GDFT}$ functional in Eq.~\eqref{eq:EHxc_def} we have access to $\chi$, thus to the force constants in Eq.~\eqref{eq:C_def} with both excitonic and dynamical effects.

\paragraph{Variational formulation---}
Our formulation belongs to the set of functional approximations with nonlocal exchange effects covered in Ref.~\onlinecite{Giovanni_2024}. 
Now we take advantage of the partially-screen partially-screen formulation of the electronic response of Ref.~\onlinecite{Giovanni_2024} to include excitonic and dynamical corrections in a differential form (see SM~\cite{supp-info}). 
The force constants associated to the GDFT functional are rewritten as
\begin{multline}\label{eq:C_psps}
    C_{s\alpha,r\beta}(z) =
    V_{\GDFT}^{(s\alpha)}(\bm r_1,-z)\chi^{\mathcal{W}}(\bm r_1,\bm r_2,z)V_{\GDFT}^{(r\beta)}(\bm r_2,z)\\
    -\rho_{\GDFT}^{(s\alpha)}(\bm r_1,-z)f^{\GDFT}_{\Hxc}(\bm r_1,\bm r_2)\rho_{\GDFT}^{(r\beta)}(\bm r_2,z) ,
\end{multline}
where $\chi^{\mathcal{W}}$ is the dynamical susceptibility with ladder diagrams, defined in Eq.\ (89) of Ref.~\onlinecite{Giovanni_2024}, also used in exciton-phonon physics~\cite{Paleari_22}. $\rho_{\GDFT}^{(s\alpha)}$ is the induced density of the GDFT functional of  Eq.~\eqref{eq:EHxc_def} and
\begin{equation}
    V_{\GDFT}^{(s\alpha)}(\bm r,z) = 
V_{\ext}^{(s\alpha)}(\bm r)+
f^{\GDFT}_{\Hxc}(\bm r,\bm r_1)\rho_{\GDFT}^{(s\alpha)}(\bm r_1,z),
\end{equation}
where $f_{\Hxc}^{\GDFT} = f_{\Hxc}^{\DFT}+\Delta f_{\Hxc}$  and
$f^{\DFT}_{\Hxc}(\bm r,\bm r\p) = \partial^2
E^{\DFT}_{\Hxc}/\partial \rho(\bm r)\partial \rho(\bm r\p)$ are the GDFT and DFT local kernels, and $\Delta f_{\Hxc}{=}\partial^2 \bar{E}_{\xc}/\partial \rho(\bm r)\partial \rho(\bm r\p)$.
The reformulation in Eq.\ \eqref{eq:C_psps} ensures that an approximation on the density $\rho_{\GDFT}^{(s\alpha)}$ results in an error on the force constants that is quadratic with respect to that on $\rho_{\GDFT}^{(s\alpha)}$.
As in Ref.~\onlinecite{Calandra_10}, we approximate $\rho_{\GDFT}^{(s\alpha)}(\bm r,z) \approx \rho_{\DFT}^{(s\alpha)}(\bm r)$, where $\rho_{\DFT}$ is the static induced-density obtained from DFT.
Such an approximation corresponds to considering $\mathcal{W}{=}0$ in Eq.~\eqref{eq:EHxc_def}.
After some algebra, the force constant matrix elements read
\begin{equation}
  C_{s\alpha,r\beta}(z) = C^{\DFT}_{s\alpha,r\beta}+\Delta C_{s\alpha,r\beta}(z)+\mathcal{O}(|\rho_{\GDFT}-\rho_{\DFT}|)^2 \label{eq:CeqCdeltaC},  
\end{equation}
where $C^{\DFT}_{s\alpha,r\beta} {=} V_{\ext}^{(s\alpha)}(\bm r_1)\rho_{\DFT}^{(r\beta)}(\bm r_1)$
are the static force constants within a DFT formalism [Eq.~\eqref{eq:EHxc_def} with $\mathcal{W}{=}0$], and 
\begin{align}
\Delta C_{s\alpha,r\beta}(z)&=\tilde{V}_{\GDFT}^{(s\alpha)}(\bm r_1)\chi^{\mathcal{W}}(\bm r_1,\bm r_2,z)\tilde{V}_{\GDFT}^{(r\beta)}(\bm r_2) \nonumber\\
&-V_{\DFT}^{(s\alpha)}(\bm r_1)\chi^0_{\DFT}(\bm r_1,\bm r_2)  V_{\DFT}^{(r\beta)}(\bm r_2)\nonumber \\
&-\rho_{\DFT}^{(s\alpha)}(\bm r_1)\Delta f_{\Hxc}(\bm r_1,\bm r_2)\rho_{\DFT}^{(r\beta)}(\bm r_2)\label{eq:Delta_C}
\end{align}
is a dynamical correction that includes excitonic effects though $\chi^\mathcal{W}$.
$\chi^0_{\DFT}$ is the static bare susceptibility evaluated with   DFT orbitals, energies and occupations (see SM~\cite{supp-info}), while 
\begin{equation}
    \tilde{V}_{\GDFT}^{(s\alpha)}(\bm r) = 
V_{\ext}^{(s\alpha)}(\bm r)+
f^{\GDFT}_{\Hxc}(\bm r,\bm r_1)\rho_{\DFT}^{(s\alpha)}(\bm r_1) \label{eq:vextHxc}
\end{equation}
is the GDFT vertex evaluated at the DFT  density and 
$V_{\DFT}^{(s\alpha)}(\bm r) = 
V_{\ext}^{(s\alpha)}(\bm r)+
f^{\DFT}_{\Hxc}(\bm r,\bm r_1)\rho_{\DFT}^{(s\alpha)}(\bm r_1)$.
$\Delta C_{s\alpha,r\beta}$ accounts for the deviation from the  DFT due to excitonic and dynamical effects.
Assuming that $\chi^{\mathcal{W}}$ and $\chi^0_{\mathrm{DFT}}$ are different only for few electronic bands \cite{Calandra_10,Berges_23,Giovanni_2024},  the differential nature of Eq.~\eqref{eq:Delta_C} allows to compute $\Delta C_{s\alpha,r\beta}$ via low-energy models, boosting the efficiency of the numerical implementations.
The local term $\rho_{\DFT}\Delta f_{\Hxc}\rho_{\DFT}$ in Eq.~\eqref{eq:Delta_C} can be trivially evaluated.

{Note that, if $\Delta f_{\Hxc}$ is set to zero in Eqs.\ \eqref{eq:Delta_C}-\eqref{eq:vextHxc}, $\tilde{V}_{\GDFT}{=}V_{\DFT}$ and the correction of the force constant matrix $\Delta C(z)$ consist in the difference between  $\chi^{\mathcal{W}}(z)$  and $\chi^0_{\DFT}$. From such difference, it is explicit how $\Delta C (z)$ contains both dynamical and excitonic effects missed in standard DFT.}

\paragraph{Computational approach---}
The static DFT dynamical matrix $C^{\DFT}_{s\alpha,r\beta}$ and density $\rho_{\DFT}^{(r\beta)}(\bm r)$ are obtained with the Quantum-ESPRESSO package~\citep{QE_2020}. The excitonic and/or dynamical corrections $\Delta C(z)$ in Eq.~\eqref{eq:Delta_C} are computed using the interacting, $\pi$-band, tight-binding approach of Ref.\ \onlinecite{albertoTB}, where the
 screened interaction $\mathcal{W}$ is self-consistently evaluated with the band structure~\cite{albertoTB}.
The parameters of the model are fitted on ab initio calculations. Reducing the Hilbert space, we neglect the contribution to Eq.~\eqref{eq:Delta_C} of the {$\bar{E}_{\xc}$ contribution}.
We calculate the phonon frequency and full-width half maximum (FWHM) from the peak position and width of the spectral function
\begin{equation}
A(\omega)={-}\sum_{s\alpha}{\rm Im}\left[\frac{\omega}{\pi} G_{s\alpha,s\alpha}(\omega)\right] 
\end{equation}
(details in SM \cite{supp-info}).

\begin{figure}[h!]
\centering
{\includegraphics[width=0.9\columnwidth]{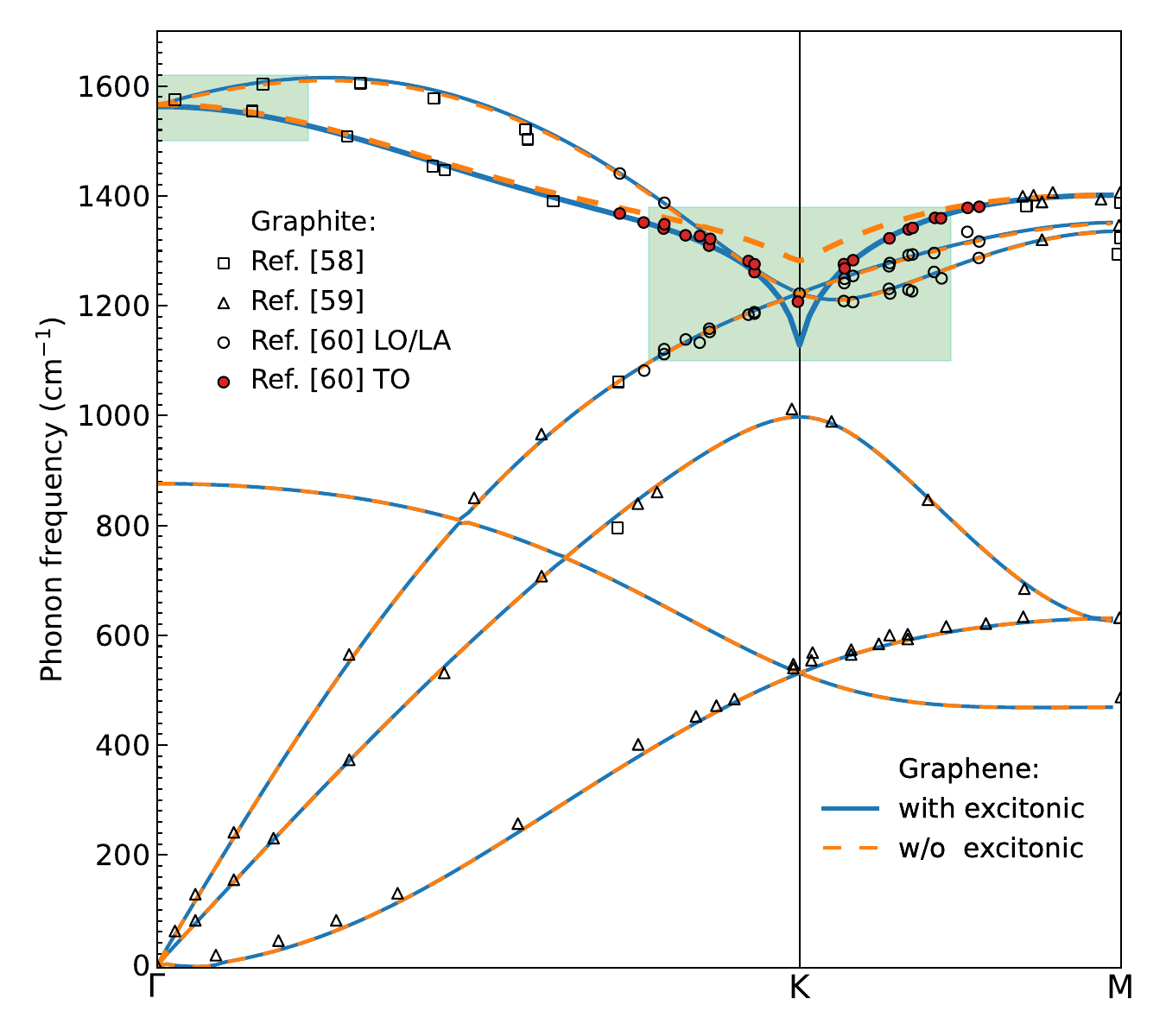}}
\caption{Theoretical static phonon dispersion of graphene (lines) compared with experimental data of graphite obtained from Refs.\ \onlinecite{Maultzsch_04} (squares), \onlinecite{Maultzsch_07} (triangles) and \onlinecite{Gruneis_09} (circles).
Red circles: data obtained for the TO branch \cite{Gruneis_09}. Orange line: results obtained by DFT calculations [$\mathcal{W} {=} 0$ in Eq.~\eqref{eq:EHxc_def}] on graphene~\cite{Piscanec_2004,Attaccalite_2009,Venezuela_11}. Blue line: GDFT result of this work containing excitonic corrections [$\mathcal{W} {\neq} 0$ in Eq.~\eqref{eq:EHxc_def}]. The shaded regions are detailed in Fig.~\ref{fig:static}. }
\label{fig:all}
\end{figure}

\textit{Application to neutral graphene---} In Fig.\ \ref{fig:all}, we show the static phonon dispersion of graphene obtained by setting $\w{=}0$ in  Eq.\ \eqref{eq:CeqCdeltaC}. We do so both within the GDFT including excitonic effects [$\mathcal{W}{\neq} 0$ in Eq.~\eqref{eq:EHxc_def}] and within DFT, neglecting excitonic effects  [$\mathcal{W}{=}0$ in Eq.~\eqref{eq:EHxc_def}].
Our calculations reveal an enhanced redshift of the TO phonon energy near the K point in graphene with respect to graphite, consistent with the expected dimensionality effects on the screening of $W$. Instead, at $\Gamma$ we find no dimensionality-related excitonic effect on the scale of Fig.\ \ref{fig:all}, as discussed below. The overall excellent agreement of our phonon dispersions with experimental data ~\cite{Maultzsch_04,Maultzsch_07,Gruneis_09} in whole BZ validates, a posteriori, the disregard of {the $\bar{E}_{\xc}$ term in Eq.\eqref{eq:EHxc_def}}.

\begin{figure}[h!]
\centering
\includegraphics[width=0.9\linewidth]{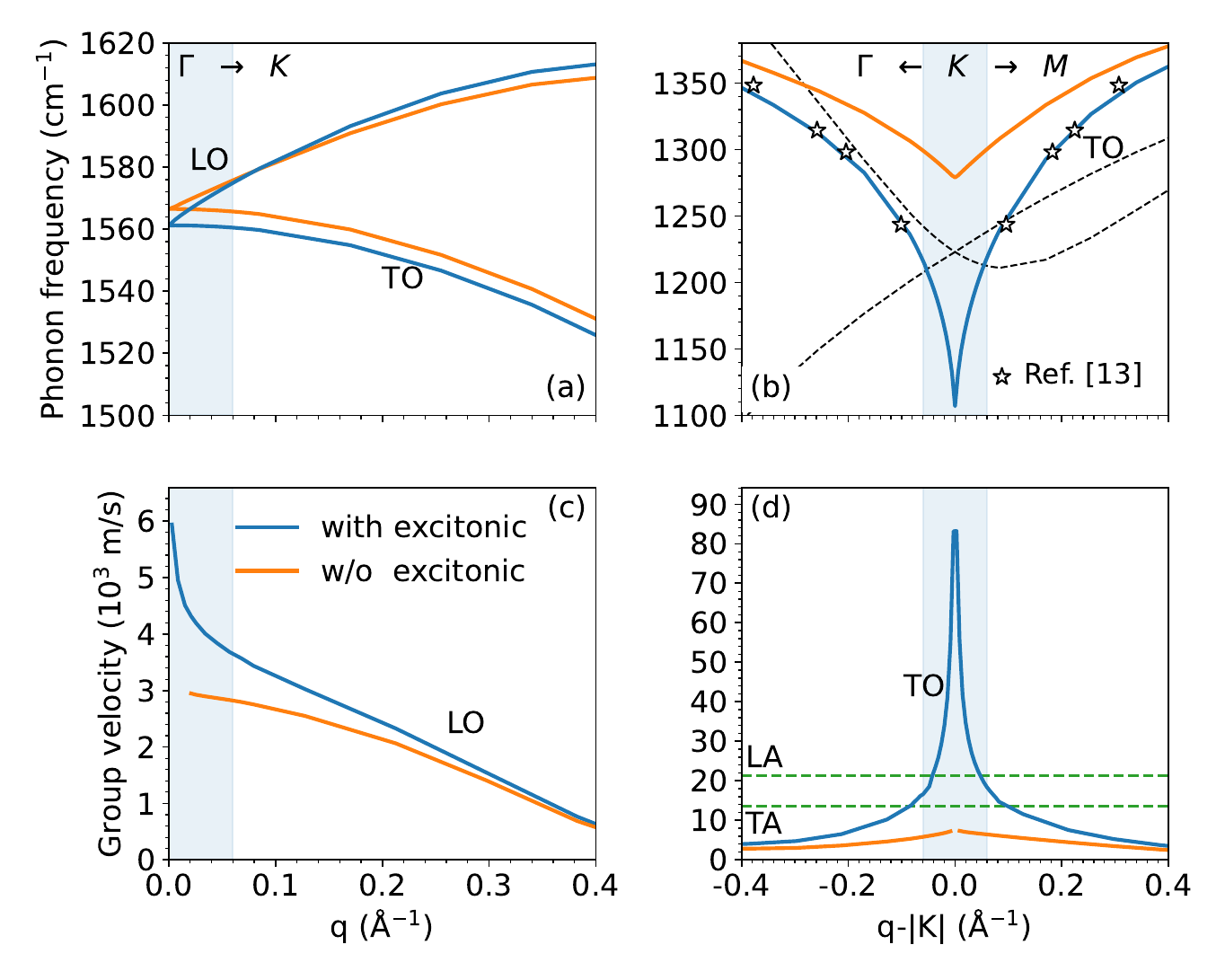}
\caption{Dispersion of the static phonons of graphene near  $\Gamma$ (a) and K  (b), and their group velocities [(c) and (d) respectively], at $T=70$ K. Orange: results from DFT (no excitonic effects) in the GGA. Blue: calculations including static excitonic effects. Green dashed lines: sound velocity of acoustic branches (LA, TA), reported for comparison. Blue shaded areas: momentum domain of Fig.~\ref{fig:dynamic} and \ref{fig:doped_Kappa} where electron and phonon excitations are close to resonance, where dynamical effects are important.
{In panel (b), we show for completeness the LO and LA phonon branches obtained from DFT with black dashed lines. Stars represent Raman experimental data from Ref.~\cite{Venanzi_2023}, red-shifted by $20$ cm$^{-1}$, which correspond to the difference between the G peak ($\sim 1580$ cm$^{-1}$~\cite{Yan_2007,Froehlicher_15,Sonntag_23}) and our phonon frequency at $\Gamma$ ($1560$ cm$^{-1}$).}
}
\label{fig:static}
\end{figure}

In Fig.~\ref{fig:static}, we show the dispersion of the static optical phonons of freestanding graphene around $\Gamma$ [LO and TO, panel (a)] and K [TO, panel (b)], and their group velocity $v_{\mathrm{G}}(q) {=} \partial \omega(q)/\partial q$ [panel (c-d)]. 
The DFT phonon dispersions [i.e.\  without excitonic effects setting $\mathcal{W}{=}0$ in Eq.\ \eqref{eq:EHxc_def}] of the LO mode around $\Gamma$ and the TO mode around K show two Kohn anomalies with slopes $v_{\textrm{G}}^{\mathrm{LO}}(\Gamma) {=} 2.9 {\times} 10^3$ m/s and $v_{\textrm{G}}^{\mathrm{TO}}(\mathrm{K}) {=} 7.3 {\times} 10^3$ m/s ~\cite{Piscanec_2004}. Excitonic effects enter as follows.
Around $\Gamma$, the slope of the LO phonon dispersion is slightly increased, with a small divergence of the Fermi velocity very close to $q{=}0$.
The TO phonon dispersion around $\Gamma$ is instead not affected by excitonic corrections, except for a minor shift of $4$ cm$^{-1}$, which is also found in the LO branch.
{This is realistically due to the neglecting of the $\bar{E}_{\xc}$ term in Eq.~\eqref{eq:EHxc_def}.}
{The DFT phonon frequency at $q=\Gamma$ ($1560$ cm$^{-1}$) is in accordance with the Raman G peak position $\sim 1580$ cm$^{-1}$~\cite{Yan_2007,Froehlicher_15,Sonntag_23}, as already noticed in the literature~\cite{Piscanec_2004,Attaccalite_2009}.
A small shift of the optical band ($20$ cm$^{-1}$), independent of momentum transfer, can be expected from small deviations in the lattice parameter~\cite{Piscanec_2004}, choice of xc functional or temperature effects.
Excitonic effects correctly do not alter this good comparison.
}
Excitonic effects have instead a great impact on the TO phonon dispersion around K, with a maximum phonon frequency shift of 173 cm$^{-1}$.
The group velocity asymptotically increases while reaching the Dirac point, with a tenfold enhancement with respect to DFT velocities.
{Our results are in excellent agreement with Raman measures of the 2D peak from Ref.~\cite{Venanzi_2023}, after the rigid shift of the experimental data by the $20$ cm$^{-1}$ previously discussed.}

\begin{figure}[h]
\centering
{\includegraphics[width=0.9\linewidth]{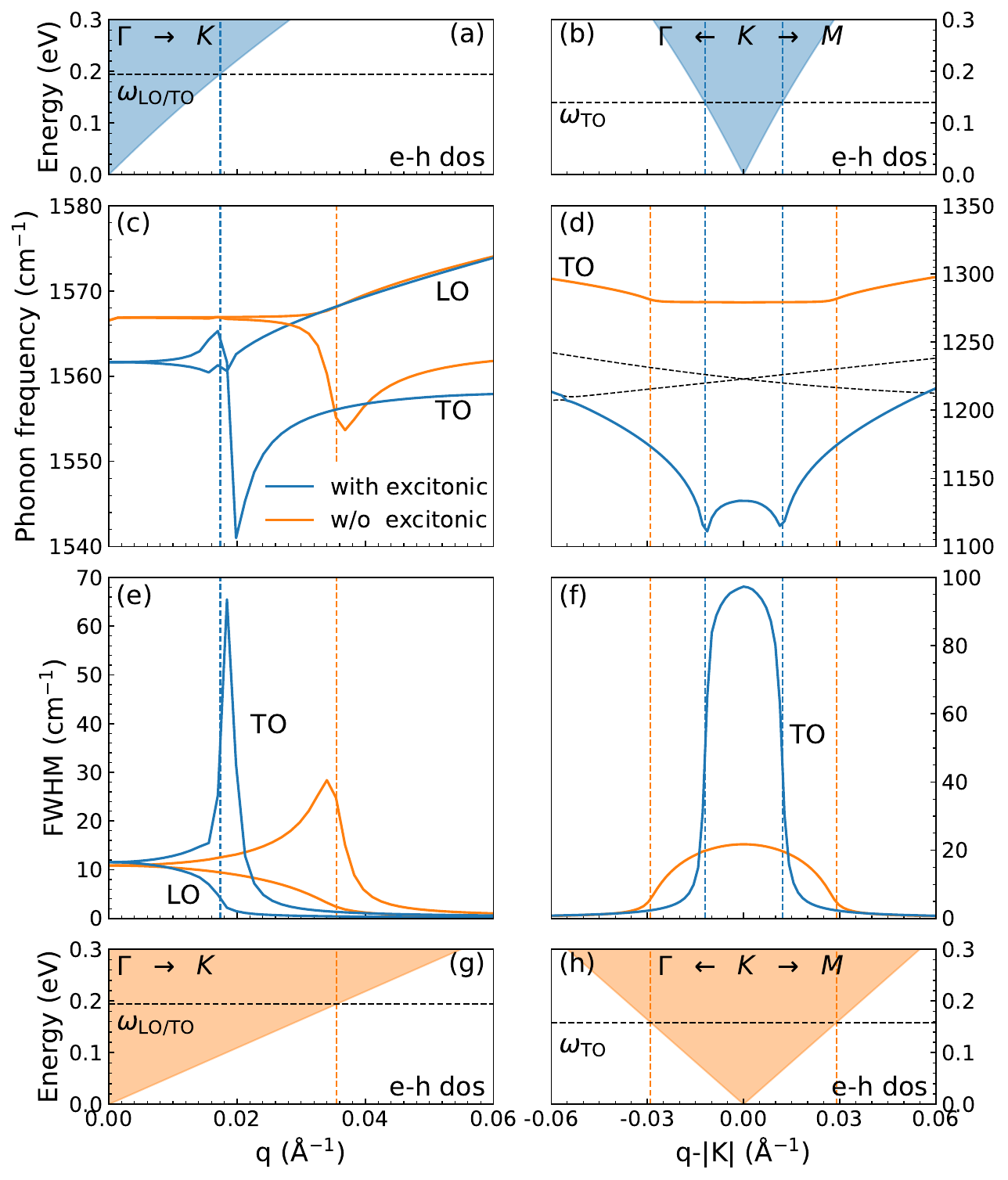}}
\caption{Top and bottom panels [(a)-(b)-(g)-(h)]: {regions where the} joint density of states {is different from zero,} with GDFT (blue) and DFT (orange) for neutral graphene near  $\Gamma$ (a)-(g) and near  K (b)-(h). Color-filled regions are where e-h excitations are possible. Vertical dashed lines delimit the regions where phonons are in resonance with electronic excitations . 
Middle panels (c)-(d): dispersion of dynamical TO and LO phonons. Middle panels (e)-(f): full-width half maximum (FWHM) for the same branches.
{The slopes of the edges of the blue/orange lines are determined by the electron group velocity, as discussed in the SI~\cite{supp-info}. In panel (d), we show for completeness the LO and LA phonon branches obtained from DFT with black dashed lines.}
}
\label{fig:dynamic}
\end{figure}

In Fig.~\ref{fig:dynamic} we analyze the features originating from the dynamical dependence of the phonon response, which is governed by {regions where} the joint density of states, i.e.\ the  energy/momentum phase space presenting electron-hole excitations, {is different from zero}.
Nonlocal exchange effects change the phase space due to an increase of the Fermi velocity{~\cite{Trevisanutto_2008,Attaccalite_2009,Guandalini_2024,albertoTB,Elias_11,Sonntag_2018}}, affecting the resonant momentum domain between lattice and electronic excitations. In Fig.\ \ref{fig:dynamic}({c})-(d), we show the dynamical phonon dispersion obtained by setting $\w$ equal to the phonon frequency in  Eq.\ \eqref{eq:CeqCdeltaC}. Once again, we do so both within the GDFT including excitonic effects [$\mathcal{W}{\neq} 0$ in Eq.~\eqref{eq:EHxc_def}] and in DFT,  neglecting excitonic effects  [$\mathcal{W}{=}0$ in Eq.~\eqref{eq:EHxc_def}]. The dynamical effects show up in freestanding graphene as sudden changes of the phonon dispersion and of the full-width half maximum (FWHM) (Fig.\ \ref{fig:dynamic}(e)-(f)). We find a large excitonic renormalization around K of ${\sim}150$ $ \text{cm}^{-1}$ also for the dynamical phonons frequency. Such an enhancement is far more evident for the FWHM, which is $\sim 5$ larger than the DFT result, increasing from ${\sim}20$ cm$^{-1}$ to ${\sim} 100$ cm$^{-1}$. 
%Phonon frequencies around $\Gamma$ are less influenced by excitonic corrections.
The main effects around $\Gamma$ are a reduction of the resonant region and an increase of the TO phonon frequency on the left side of the resonant separation, and the FWHM is extremely similar to the DFT one at zone center{, compatible with the Raman experimental value of the G peak in the range of $12$-$15$ cm$^{-1}$~\cite{Yan_2007,Froehlicher_15,Sonntag_23}.}

%{To distinguish between self-energy and ladder/excitonic corrections on the phonon frequencies and lifetimes, we refer to Fig. 1 of the SM~\cite{supp-info}.}

We rationalize the above results in terms of electron-phonon couplings. We crucially exploit the exact generalized Fermi golden rule derived in Ref.\ \cite{Giovanni_2024}, 
which allows to determine the exact FWHM in terms of dressed electron-phonon couplings as
\begin{align}
\textrm{FWHM}\Big{|}_{\indxlt}=\frac{Ad_{\indxlt}}{2\hbar}   \frac{|g^{\indxlt}|^2\hbar\omega_{\indxlt}}{\hbar v_{\mathrm{g}}\hbar v_{\phi}},
\label{eq:FWHM}
\end{align}
where $A$ is the area of the cell, $d_{\indxlt}=2/1$ is the Dirac cone degeneracy, $v_{\mathrm{g}}$ and $v_{\phi}$ are respectively the electronic group and phase velocities, and $g^{\indxlt}$ is the electron-phonon matrix element between the phonon of frequency $\omega_{\indxlt}$ and the $\pi$ and $\pi^*$ bands. 
{A similar formula has been used in Ref.~\cite{Yan_2007} to analyze the Raman G peak.} 

The elements of Eq.\ \eqref{eq:FWHM} are evaluated at the electronic momentum resonant with $\omega_{\indxlt}$ (see SM~\cite{supp-info}). Comparing the FWHM computed with GDFT [$\mathcal{W}{\neq}0$ in Eq.\ \eqref{eq:EHxc_def}] and with DFT [$\mathcal{W}{=}0$ in Eq.\ \eqref{eq:EHxc_def})], we deduce the enhancements of the electron-phonon couplings due to excitonic effects. At K, using Eq.\ \eqref{eq:FWHM} and the values reported in the SM, we find $|g^{\mathrm{K},\GDFT}|^2{\sim} 26|g^{\mathrm{K},\DFT}|^2$.
{This strong enhancement of $g^{\text{K}}$ is in accordance with the analytical prediction of Refs.~\cite{Basko_2008,Basko_2009}, where a logarithmic divergence of the electron-phonon coupling at K is expected at $T{=}0K$ for undoped graphene.}
At $\Gamma$ instead we have $|g^{\Gamma,\GDFT}|^2{\sim} 5|g^{\Gamma,\DFT}|^2$. Notably, the enhancement of the FWHM at K is not compensated by the increase of the electronic group/phase velocities, while this compensation happens at $\Gamma$ {(see the SM~\cite{supp-info})}. In fact, the overall insensitiveness of phonon features to excitonic effects at zone center is due to the gauge field nature of the electron-phonon coupling. Indeed, as shown in the SM~\cite{supp-info}, the effect of an atomic displacement is to simply shift the origin of the band structure from the K point, even in presence of electron-electron interactions. 
As a consequence, the electron-phonon coupling reads $|g^{\Gamma}|{=}\frac{\beta}{b_0^2}v_{\textrm{g}}\sqrt{\frac{\hbar}{2M\omega_{\Gamma}}}$, where $b_0$ is the interatomic distance and $\beta$ is an adimensional coupling parameter, untouched by the electron-electron interaction (see SM for more details). It follows that the FWHM at zone center is proportional to $v_{\textrm{g}}/v_{\phi}$, which is weakly dependent on excitonic effects. Notice that this result can be seen as a very stringent and non trivial test for the self-consistency of the theory. For example, as shown numerically in the {end matter}, the gauge field protection of the FWHM is broken if only a small number of ladder diagrams are considered in the determination of the phonon's Green functions, and is recovered only if the full series is computed. Similarly, but in a less stringent  fashion, the same arguments explain the different behaviors of the phonon frequencies at $\Gamma$ and K.

As further analysis, we append in the end matter of this letter our study of the behavior of the FWHM on doped graphene.

\textit{Outlook---}
We showed how to calculate phonon dispersions and lifetimes with excitonic effects included in the density-density response function within a generalized {Kohn-Sham} framework.
We found that excitonic interactions have huge effects on the optical zone border phonons of graphene near K, where conducting electrons are scattered by the phonons.
Phonons near $\Gamma$ are instead poorly influenced by excitonic effects, due to a gauge field protection which causes an opposite renormalization of phonon frequencies and lifetimes due to the increase of both the group velocity and electron-phonon coupling.
Combining previous theoretical estimations of the resistivity~\cite{Park_2014_2} with the strong increase of electron-phonon scattering observed here, we can infer that the Zone border phonons near K are likely to be the dominant source of intrinsic resistivity in graphene at room temperature.
{Excitonic effects may be relevant also in the description of electron-phonon coupling in high-correlated and/or superconducting materials~\cite{yin2013correlation,calandra2015universal} .}
{Being an effect mostly driven by dimensionality, we expect excitonic effects to be as relevant on the response of low dimensional system such as carbon nanotubes or carbon atom chains, where exchange effects are known to be crucial also for electronic structure calculations \cite{romanin2021dominant,Villani2024}. }

\paragraph{Acknowledgments---}
This project has received funding from the European Research Council (ERC) under the European Union’s Horizon 2020 research and innovation programe (MORE-TEM ERC-SYN project, grant agreement No 951215). We thank S. Laricchia for useful discussions. We dedicate the findings of this work to the memory of N. Bonini, who
brought to our attention the possibility to include excitonic effects in
phonon calculations within density-density response theory.\\

\bibliography{bibliography}
\clearpage
\appendix
\section{End matter}
\begin{figure}[h!]
\centering
\includegraphics[width=0.9\columnwidth,]{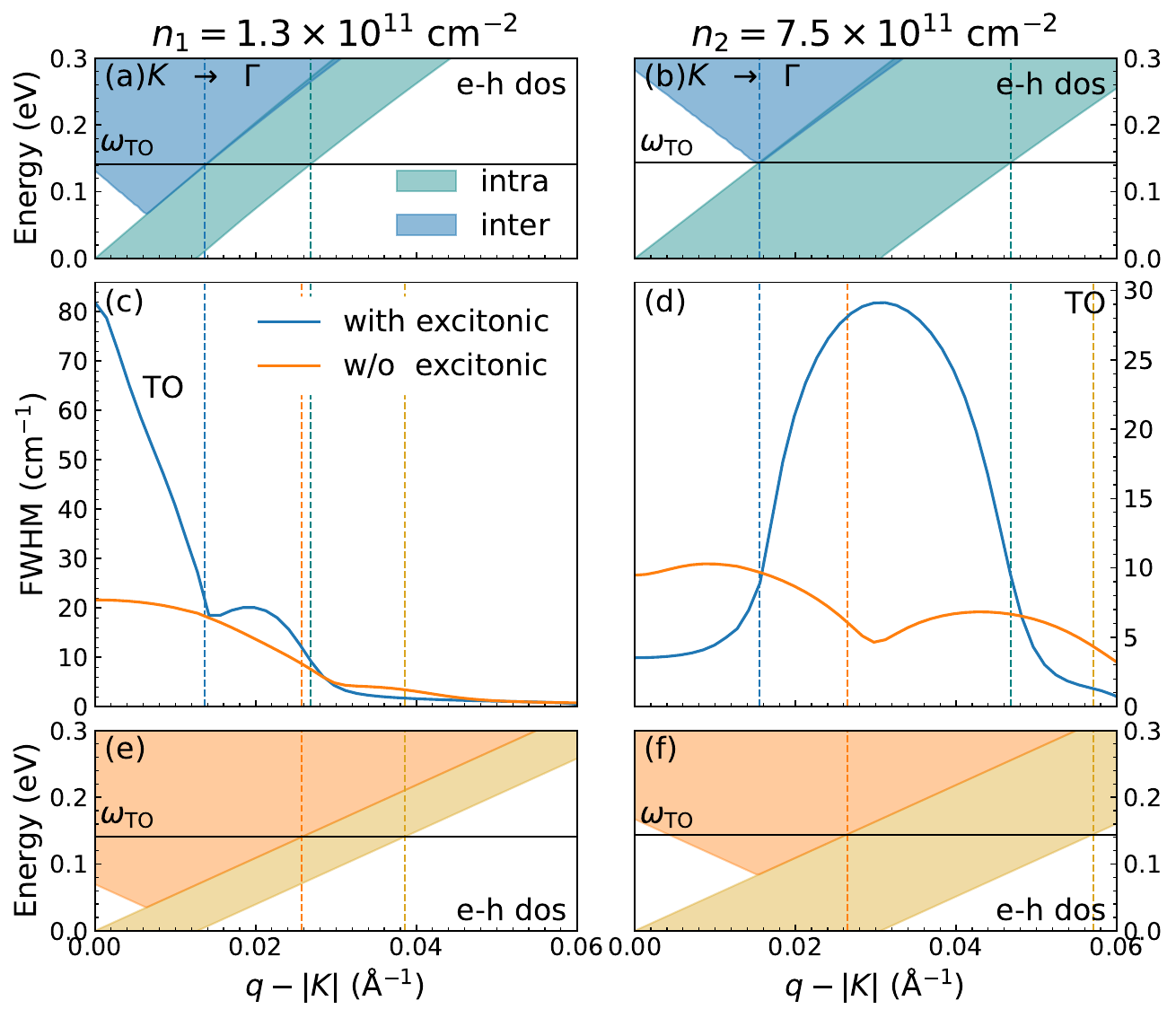}
\caption{(a)-(b): {regions where the } joint density of states {is different from zero} in GDFT (blue), and in DFT (orange). 
Intra and inter-band transitions are represented with different colors. (c)-(d): TO phonon FWHM near K. (e)-(f): joint density of states with DFT. Left and right columns refers to the different doping levels indicated in figure.
{The slopes of the edges of the blue/orange lines are given by the electron group velocity, discussed more in detail in the supplemental information~\cite{supp-info}.}
}
\label{fig:doped_Kappa}
\end{figure}

\paragraph{Doping effects ---}We present in Fig.~\ref{fig:doped_Kappa} the effect of doping in the FWHM and {regions where the} joint density of states {is different from zero} near K for two different doping levels, $n_1 {=} 1.3 {\times} 10^{11}$ cm$^{-2}$ and $n_2 {=} 7.5 {\times} 10^{11}$ cm$^{-2}$  which are upper limits for the densities relevant for recent transport measurements \cite{PhysRevB.107.075420}. In the SM, we present the same study near $\Gamma$. Fixed doping level corresponds to different Fermi energies at different levels of theories, i.e. $\left[E_{\textrm{F}}(n_1)\right]_{\GDFT}{=}0.066$eV, $\left[E_{\textrm{F}}(n_1)\right]_{\textrm{DFT}}{=}0.035$eV, and $\left[E_{\textrm{F}}(n_2)\right]_{\GDFT}{=}0.14$eV, $\left[E_{\textrm{F}}(n_2)\right]_{\textrm{DFT}}{=}0.08$eV. The nature of the peaks is qualitatively the same, a part for an additional peak originated from the decay of phonons into intraband e-h pairs. Quantitatively, at $n_1$ the FWHM maintains a large excitonic renormalization, even if slightly reduced due to screening effects of the Dirac electrons. At $n_2$ the interband peak is quenched as the interband electronic transitions are no more in resonance with the phonons due to the partial filling of the upper Dirac cone. This condition arises when $2E_{\F} > \omega_{\mathrm{TO}}$.
The intraband peak is instead similarly enhanced at both doping levels. For the electron-phonon couplings, at $n_1$ we find $|g^{\mathrm{K},\GDFT}|^2{\sim }15|g^{\mathrm{K},\DFT}|^2$ and $|g^{\Gamma,\GDFT}|^2{\sim} 4|g^{\Gamma,\DFT}|^2$.

{
\paragraph{Splitting between self-energy and ladder corrections, and convergence with the number of ladder diagrams ---}
In Eq.~\eqref{eq:Delta_C}, the difference between $\chi^{\mathcal{W}}$ and $\chi^{0}_{\DFT}$ has two contributions: (i) the different electronic band structure due to $\mathcal{W} \neq 0$ in Eq.~\eqref{eq:EHxc_def} (self-energy corrections); (ii) the presence of electron-hole interaction diagrams in $\chi^{\mathcal{W}}$ (ladder corrections).
In order to distinguish the two contributions, we plot in Fig.\ \ref{fig:ladder} the phonon frequencies (a)-(d) and FWHM (e)-(f) of phonons around $\Gamma$ (left panels) and K (right panels), obtained via DFT calculations ($\mathcal{W} = 0$) or starting from the band structure computed with $\mathcal{W} \neq 0$ and then including an increasing number $N$ of ladder diagrams in the series expansion of the electron-hole propagator $L^{\text{W}}$ used to compute the phonon response (defined such that $\chi^{\mathcal{W}}(\bm r_1,\bm r_2,z) = L^{\text{W}}(\bm r_1,\bm r_1,\bm r_2,\bm r_2,z)$ \cite{supp-info,Giovanni_2024} ).
%  and highlight the importance to include the infinite series of ladder diagrams in the calculation of $\chi^{\mathcal{W}}$
%As a further analysis, we present in Fig.\ \ref{fig:ladder} the phonon dispersions and lifetimes with growing excitonic effects. We do so by increasing the number $N$ of ladder diagrams included in the particle-hole propagator $L^{\text{W}}$ used to compute the phonon response, defined such that $L^{\text{W}}(\bm r_1,\bm r_2,\bm r_3,\bm r_4,z) {=}\chi^{\mathcal{W}}(\bm r_1,\bm r_2,z) $ (more details in SI \cite{supp-info} and \cite{Giovanni_2024}).

\begin{figure}[h!]
\centering
\includegraphics[width=0.9\linewidth]{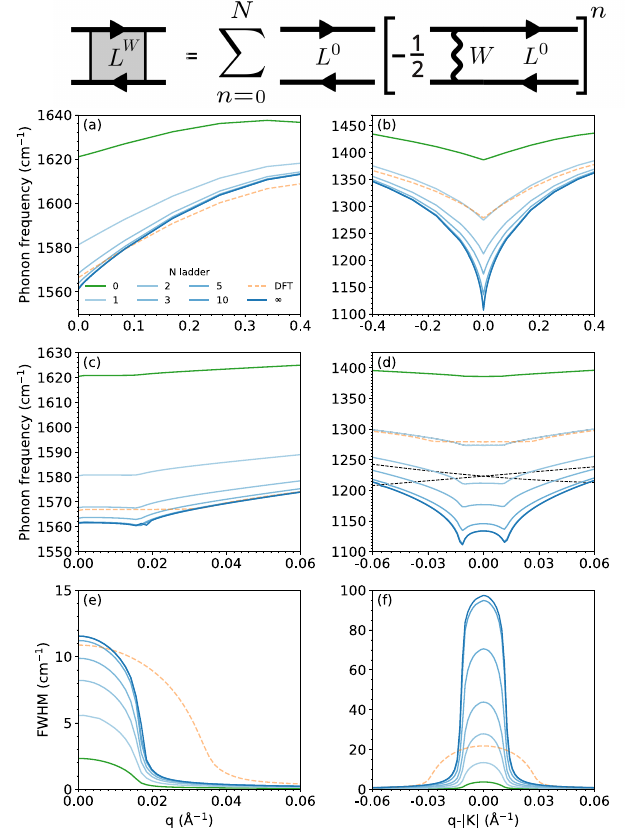}

\caption{{Impact of the number of electronic ladder diagrams on the phonon frequency and FWHM at $\Gamma$ (left panels) and K (right panels). Starting from the DFT results (orange), self-energy effects (green) blue-shift the phonon frequencies and reduce the FWHM, as opposite with ladder corrections. For a discussion of the variation of the linewidth with the ladder, see \cite{supp-info}. In panel (d), we show for completeness the LO and LA phonon branches obtained from DFT with black dashed lines.}}
\label{fig:ladder}
\end{figure}

The difference between the DFT and the $N=0$ results is only due the self-energy corrections, i.e. to the role of $\mathcal{W}$ in determining the electronic bands (discussed thoroughly in Ref.~\cite{albertoTB}). Due to the large increase of the electron velocity due to the self-energy, the $N=0$ phonon frequencies are strongly blue-shifted and the FWHM is reduced by $\sim 5$ with respect to DFT. The role of vertex corrections is to completely revert this conclusion: at K, the $N=\infty$ frequencies are strongly red-shifted and the FWHM is enhanced by $\sim 5$ with respect to DFT, while at $\Gamma$ frequencies and FWHM are almost coincident with DFT results, as expected from the gauge-field protection. The same trend is observed also for the value of the electron-phonon coupling at K. The results at $\Gamma$ underline the importance of self-consistent theories able to satisfy exact properties, such as sum rules. In conclusion, self-energy effects and ladder corrections, both of huge impact on the observables, have different signs, as already observed for the graphene $\pi$ plasmon dispersion~\cite{guandalini_2023}, and they compensate at $\Gamma$ while excitonic effects are dominant near K, thus must be employed together.

% We note the infinite series is required to recover the DFT phonon frequency and FWHM at $q=\Gamma$,  gauge-field protected.
% This underlines the importance of self-consistent theories able to satisfy exact properties, such as sum rules etc.
% In addition, $N\ge 10$ is required also to reproduce phonon properties around the K point.
%These results demonstrates that the renormalization of the Kohn anomalies are excitonic effects, i.e. provided by the infinite resummation of ladder diagrams.

%The $N{=}\infty$ lines instead consist in the fully interacting phonon response where ladder diagrams are summed until self-consistency.
%By comparing the $N{=}0$, $N{=}\infty$ and DFT results, we show how self-energy corrections blue-shift the phonon frequencies and reduce the FWHM with respect to DFT, due to the increase of the electronic velocity, while ladder corrections red-shift the phonon frequencies and increase the FWHM. 
%This result highlights the excitonic character of the renormalization of the  Kohn anomaly of graphene. In fact, such an effect can not be described only with a renormalization of the electronic bands due to exchange effects.
}
\end{document}

\title{Excitonic effects in phonons: reshaping the graphene Kohn anomalies and lifetimes\\
Supplemental material}

\author{Alberto Guandalini}
\email{alberto.guandalini@uniroma1.it}
\affiliation{Dipartimento di Fisica, Universit\`a di Roma La Sapienza, Piazzale Aldo Moro 5, I-00185 Roma, Italy}

\author{Francesco Macheda}
\affiliation{Dipartimento di Fisica, Universit\`a di Roma La Sapienza, Piazzale Aldo Moro 5, I-00185 Roma, Italy}

\author{Giovanni Caldarelli}
\affiliation{Dipartimento di Fisica, Universit\`a di Roma La Sapienza, Piazzale Aldo Moro 5, I-00185 Roma, Italy}

\author{Francesco Mauri}
\affiliation{Dipartimento di Fisica, Universit\`a di Roma La Sapienza, Piazzale Aldo Moro 5, I-00185 Roma, Italy}

\maketitle

\section{Additional results}
%\subsection{Convergence with respect to the number of ladder diagrams}

\begin{comment}
\begin{figure}[h!]
\includegraphics[width=\linewidth/3]{finite_sum.png}\\
\includegraphics[width=\linewidth/2]{phonons_lad.pdf}

\caption{Impact of the number of electronic ladder diagrams on the phonon frequency and FWHM at $\Gamma$ (left panels) and K (right panels). Starting from the DFT results (orange), self-energy effects (green) blue-shift the phonon frequencies and reduce the FWHM, as opposite with ladder corrections. For a discussion of the variation of the linewidth with the ladder see Sec. \ref{sec:ladderlinewidth}. }
\label{fig:temp}
\end{figure}
\end{comment}

\subsection{Phonons and lifetimes at room temperature}
\begin{figure}[h!]
\includegraphics[width=\linewidth/2]{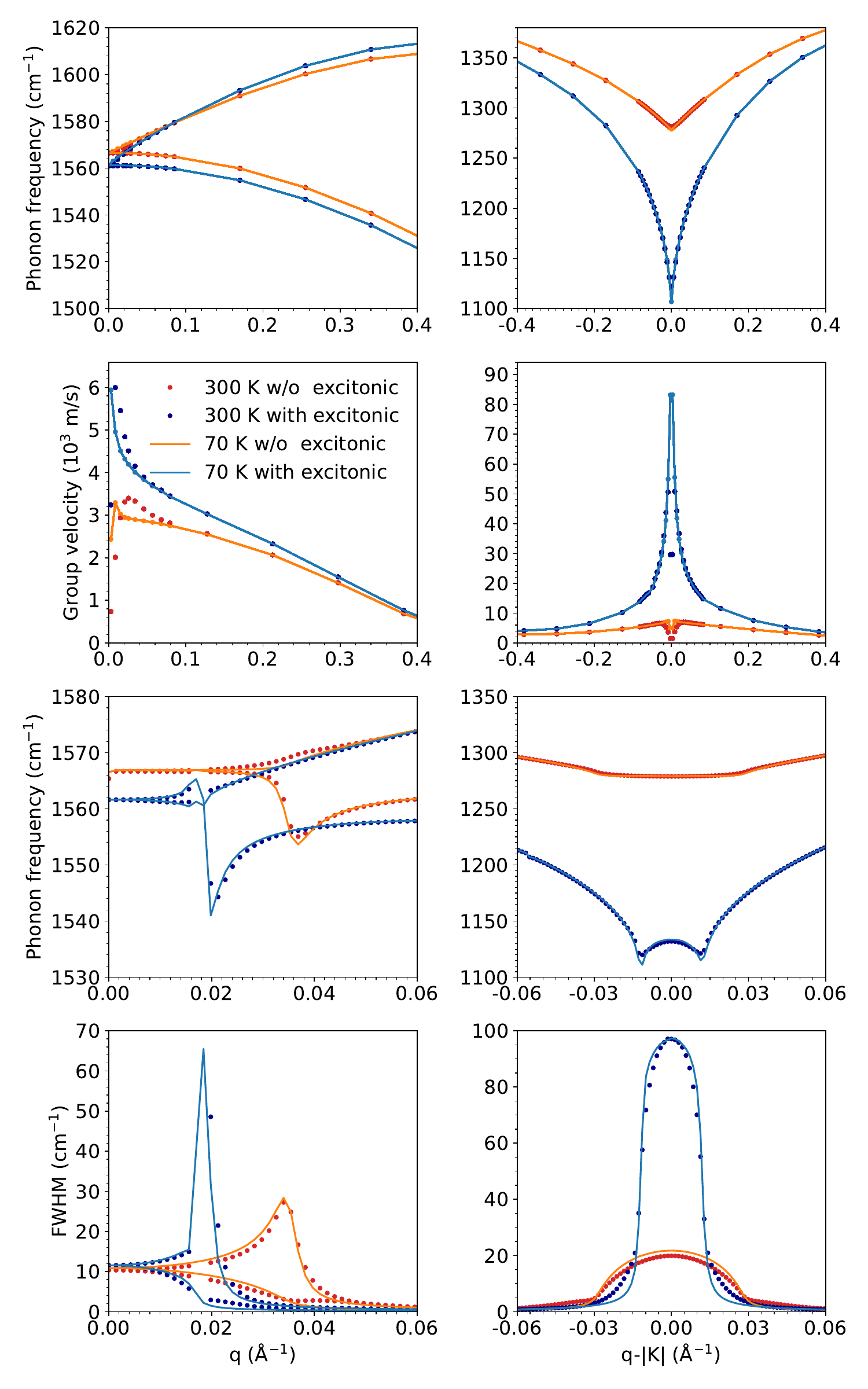}
\caption{Impact of temperature on phonon frequency, phonon velocity and FWHM at $\Gamma$ (left panels) and K (right panels).} 
\label{fig:temp}
\end{figure}
\clearpage
\subsection{Phonon lifetimes of doped graphene near $\Gamma$ at 70K}
\label{appx:Gamma_dop}

\begin{figure}[h!]
\includegraphics[width=\linewidth/2]{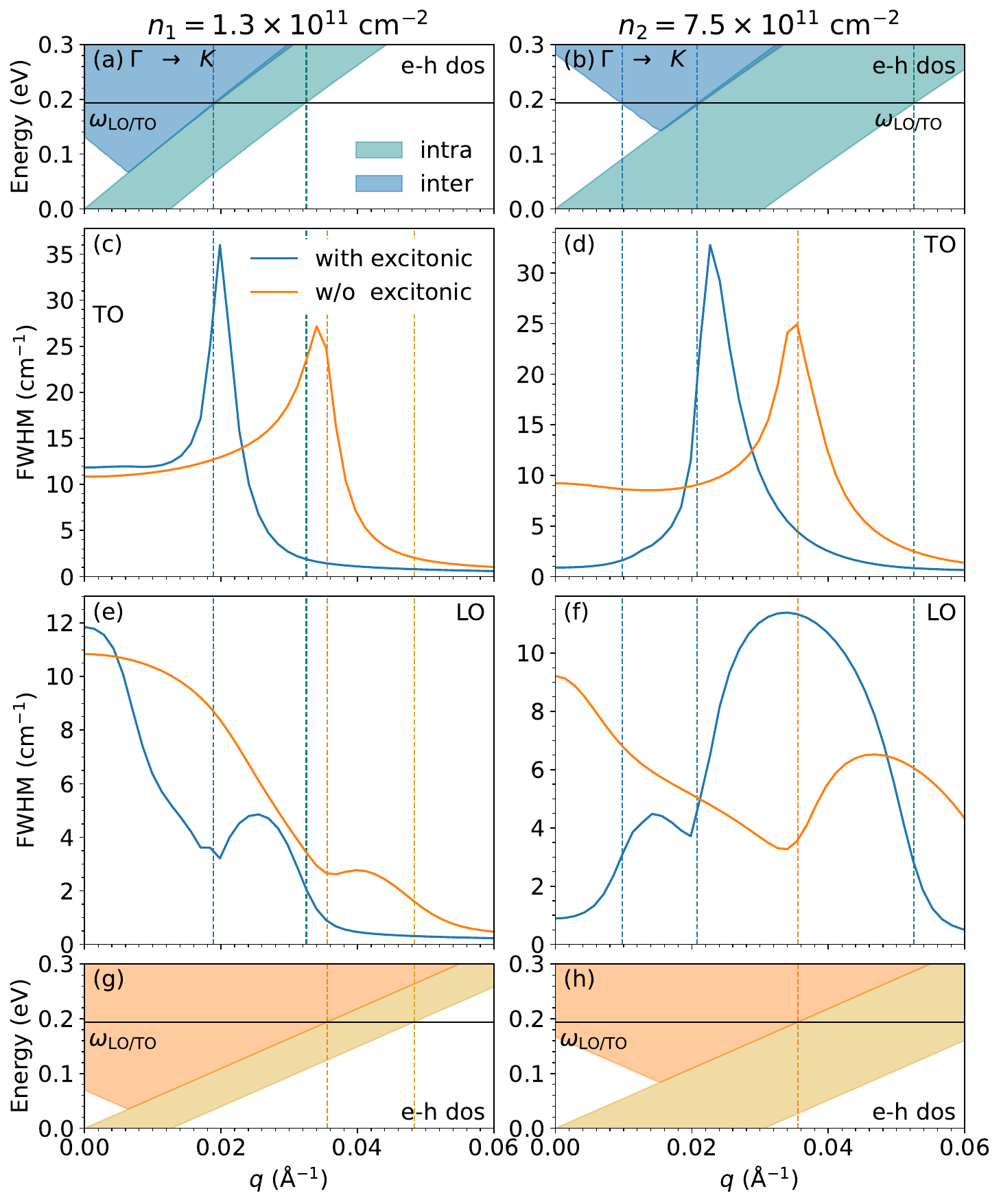}
\caption{
Joint density of states in GDFT (a)-(b) and in DFT (g)-(h).
The static phonon dispersion of the LO/TO mode (degenerate at $\Gamma$), where the dynamical matrices have been evaluated, are highlighted in black.
Intra and inter-band transitions are represented with different colors.
(c)-(d): transverse optical (TO) phonon FWHM near the $\Gamma$ point along $\Gamma$-K direction obtained with DFT in orange, while results with excitonic corrections included are in blue.
(e)-(f): same as (c)-(d) but for the longitudinal optical (LO) mode.
(a)-(c)-(e)-(g) panels show results with electron doping $n_1 = 1.3 \times 10^{11}$ cm$^{-2}$, (b)-(d)-(f)-(h) panels with $n_2 = 7.5 \times 10^{11}$ cm$^{-2}$.
Blue (orange) vertical lines separate different resonant conditions between phonons and quasi-particle (Kohn-Sham) excitations.}
\label{fig:doped_Gamma}
\end{figure}

In Fig.~\ref{fig:doped_Gamma}, we show the FWHM of the LO/TO branched near $\Gamma$ for the case of doped graphene.
nonlocal exchange effects in the band structure change the resonant regions between phonons and electron-hole excitations.
In particular, we note for $n_2$ doping at $\qph = \Gamma$ the phonons are in resonance with DFT electron-hole excitations while they are with GDFT excitations.
Excitonic effects do not produce huge renormalizations to the FWHM intensity, as already shown in the main paper for freestanding graphene.
We note the TO and LO phonons interact with different intensities with inter and intra-band electronic transitions.

\section{Sum rule and dynamical matrix}\label{appx:sum_rule}

In this Appendix, we show how to leverage the acoustic sum rule to express the dynamical matrix in terms of the induced electronic density as done in Eq.\ 2 of the main text. In linear response, the dynamical matrix is expressed as 
\begin{equation}\label{eq:D_def app}
    D_{s\alpha,r\beta}(z) =\frac{C_{s\alpha,r\beta}(z)+V_\ext^{(s\alpha,r\beta)}(\bm r_1) \rho\unpert(\bm r_1) }{\sqrt{M_sM_r}},
\end{equation}
$C_{s\alpha,r\beta}$ is defined in Eq.\ 3 of the main text, $M_l,M_r$ are the nuclear masses. $V_\ext$ is the ionic potential
\begin{equation}
    \label{v ext app}
    V_\ext(\bm r, \{\bm R_1, \dots, \bm R_M \}) = -e^2 \sum_l^{N_\text{ion}} \frac{Z_l}{|\bm r - \bm R_l|}+ \frac{e^2}{2N_{\text{el}}}\sum_{lm, l\neq m}^{N_\text{ion}} \frac{Z_lZ_m}{|\bm R_l - \bm R_m|}
\end{equation}
experienced by the electrons, which depends on the electronic coordinate $\bm r$ and parametrically on the collection of nuclear coordinates $\{\bm R_1,\dots, \bm R_M\}$.
In Eq.\ \eqref{v ext app}, $Z_l$ is the atomic number of the $l-$th ion and $e$ is the electron charge.
$V_\ext^{(s\alpha,r\beta)}$ in Eq.\ \eqref{eq:D_def app} is the derivative of $V_\ext$ in Eq.\ \eqref{v ext app} with respect to the displacements $u_{s\alpha}/u_{r\beta}$ of the nuclei $s/r$ along the Cartesian coordinates $\alpha/\beta$ \cite{Giovanni_2024}, i.e.\ 
\begin{equation}
    V_\ext^{(s\alpha,r\beta)}(\bm r) = \pdv{V_\ext\left(\bm r, \{\bm R_1\unpert{+}\bm u_1, \dots, \bm R_{N_\text{ion}}\unpert{+}\bm u_{N_\text{ion}} \}\right)}{u_{s\alpha}}{u_{r\beta}}\eval_{ \{ \bm u_1,\dots, \bm u_{N_\text{ion}}\} = \{0,\dots, 0\}},
\end{equation}
where $\{\bm R_1\unpert, \dots, \bm R_{N_\text{ion}}\unpert \}$ are the nuclear equilibrium positions. $V_\ext^{(s\alpha,r\beta)}(\bm r)$
can be expressed as (we omit the dependence on the nuclear coordinates for brevity)
\begin{equation} \label{V 2 = M + C ion}
    V_\ext^{(s\alpha,r\beta)}(\bm r) = \delta_{sr}\mathcal{M}_{s\alpha,s\beta}(\bm r) + \frac{1}{N_\text{el}}C^{\text{ion-ion}}_{s\alpha,r\beta},
\end{equation}
where $\mathcal{M}_{s\alpha,s\beta}(\bm r)$ and $C^{\text{ion-ion}}_{s\alpha,r\beta}$ read respectively
\begin{align}
  \mathcal{M}_{s\alpha,s\beta}(\bm r) & =   -e^2Z_s\frac{(r_\alpha {-}R\unpert_{s\alpha})(r_\beta {-}R\unpert_{s\beta})-\delta_{\alpha\beta}3|\bm r {-}\bm R\unpert_{s}|^2}{3|\bm r {-}\bm R\unpert_{s}|^5}  , \label{M ab}\\
    C^{\text{ion-ion}}_{s\alpha,r\beta} & = {-}e^2 Z_sZ_r\frac{(R\unpert_{s\alpha} {-}R\unpert_{r\alpha})(R\unpert_{s\beta} {-}R\unpert_{r\beta})-\delta_{\alpha\beta}3|\bm R\unpert_{s} {-}\bm R\unpert_{r}|^5}{3|\bm R\unpert_{s} {-}\bm R\unpert_{r}|^5} \nonumber \\ &{+}\delta_{sr}\sum\limits_{l}^{N_\text{ion}}e^2 Z_sZ_l\frac{(R\unpert_{s\alpha} {-}R\unpert_{l\alpha})(R\unpert_{s\beta} {-}R\unpert_{l\beta})-\delta_{\alpha\beta}3|\bm R\unpert_{s} {-}\bm R\unpert_{l}|^5}{3|\bm R\unpert_{s} {-}\bm R\unpert_{l}|^5}\label{Cion ab}.
\end{align}
The term $\mathcal{M}_{s\alpha,s\beta}(\bm r)$ in Eq.\ \eqref{M ab} depends on the electronic coordinates, while $C^{\text{ion-ion}}_{s\alpha,r\beta}$ in Eq.\ \eqref{Cion ab} only depends on ionic coordinates. The latter is divided by the number of electrons $N_\text{el}$ in Eq.\ \eqref{V 2 = M + C ion} to ensure normalization. The acoustic sum rule reads
\begin{equation}
    \label{asr}
    \sum_r^{N_\text{ion}} \sqrt{M_sM_r}D_{s\alpha,r\beta}(z)=0
\end{equation}
for every $\alpha,\beta$. Using Eq.\ \eqref{eq:D_def app} and Eq.\ \eqref{V 2 = M + C ion} in Eq.\ \eqref{asr} we get 
\begin{equation} \label{sum rule all terms}
   \sum_l^{N_\text{ion}} C_{s\alpha,l\beta}(z)+ \mathcal{M}_{s\alpha,s\beta}(\bm r_1)\rho\unpert(\bm r_1) + \sum_l^{N_\text{ion}} C^{\text{ion-ion}}_{s\alpha,l\beta} = 0.
\end{equation}
From Eq.\ \eqref{Cion ab} it can be verified directly that 
\begin{equation}
    \sum_l^{N_\text{ion}} C^{\text{ion-ion}}_{s\alpha,l\beta} = 0,
\end{equation}
so that Eq.\ \eqref{sum rule all terms} can be rewritten as 
\begin{equation}
    \label{M = sum C(z)}
    \mathcal{M}_{s\alpha,s\beta}(\bm r_1)\rho\unpert(\bm r_1) = - \sum_l^{N_\text{ion}} C_{s\alpha,l\beta}(z)
\end{equation}
which finally proves Eq.\ 2 of the main text
\begin{equation} \label{D sum rule app}
     D_{s\alpha,r\beta}(z) = \frac{C_{s\alpha,r\beta}(z)+ C^{\text{ion-ion}}_{s\alpha,r\beta} -   \sum_l C_{s\alpha,l\beta}(z)\delta_{sr}}{\sqrt{M_sM_r}}.
\end{equation}

In closing, we note that the dynamical matrix in Eq.\ \eqref{D sum rule app} depends on the first order derivative of $V_\ext$ [see Eq.\ \eqref{v ext app}] with respect to the ion displacement ---$V_{\ext}^{(s\alpha)}$ --- through $C_{s\alpha, r,\beta}(z)$ [cf.\ Eq.\ (3) main text]. Formally, $V_{\ext}^{(s\alpha)}$ contains the derivative of both terms of $V_\ext$ in Eq.\ \eqref{v ext app}, i.e.\ the electron-ion and the ion-ion interaction. However, being a constant in the electronic coordinates, the derivative of the ion-ion interaction does not produce any induced density, and can be omitted in the calculation of $V_{\ext}^{(s\alpha)}$.
%This can be verified explicitly considering in the definition of the induced density matrix (Eq.\ 6 of the main text) a constant induced potential and the fact that $\int\dd \bm r_3 L^0(\bm r_1,\bm r_2,\bm r_3,\bm r_3, z){=}0$, where $L^0$
%is the bare electron-hole propagator (Eq.\ 7  of the main text). The present discussion allows to consider in $V_{\ext}^{(s\alpha)}$ only electron-ion terms.

\section{Additive exchange-correlation term to reproduce the total energy of the uniform electron gas}\label{appx:double_counting}
In this Section we provide an explicit form of the additive term $\bar{E}_{\xc}[\rho,\mathcal{W}]$, defined in Eq. ($4$) of the main text so to reproduce the total energy of the uniform electron gas (jellium model). %To build such a functional, we have to ensure that the Hxc energy is exact in the case of a uniform electron gas.
In the local density approximation, we use the following expression:
\begin{equation} \label{Exc app}
    \bar{E}_{\xc}[\rho,\mathcal{W}] = \rho(\bm r_1)\bar{\varepsilon}_{\xc}[\mathcal{W}](\rho(\bm r_1)) \ .
\end{equation}
In Eq.\ \eqref{Exc app},  $\bar{\varepsilon}_{\xc}$ is the effective energy density of a uniform electron gas with density $\rho(\bm r)$ that cancels the Fock-like term in Eq. (4) of the main paper in the case of a uniform electron gas.
Within the local density approximation, $\bar{\varepsilon}_{\xc}$ is given by
\begin{equation}\label{eq:e_x^dc}
    \bar{\varepsilon}_{\xc}(\rho) = \frac{1}{2N_{\mathrm{el}}}
    |n^{\mathrm{HEG}}[\rho](\bm r_1,\bm r_2)|^2 \mathcal{W}(\bm r_1,\bm r_2) \ .
\end{equation}
We note that $\mathcal{W}(\bm r,\bm r\p)$ is the same screened interaction used in the nonlocal exchange contribution in the second term of Eq. ($4$) of the main paper and $n^{\mathrm{HEG}}[\rho]$ is the density matrix of the uniform electron gas with density $\rho$.

\begin{comment}
In appx. \ref{appx:periodic},
we provide $\varepsilon_{\xc}^{\mathrm{dc}}$ in Fourier space in the case of a screened interaction translationally invariant, thus $\mathcal{W}(\bm r,\bm r\p) = \mathcal{W}
(\bm r-\bm r\p)$.
\end{comment}

\section{Reformulation in terms of Feynman diagrams}\label{appx:feynman_diagrams}
\begin{figure}[h!]
\includegraphics[width=\linewidth/2]{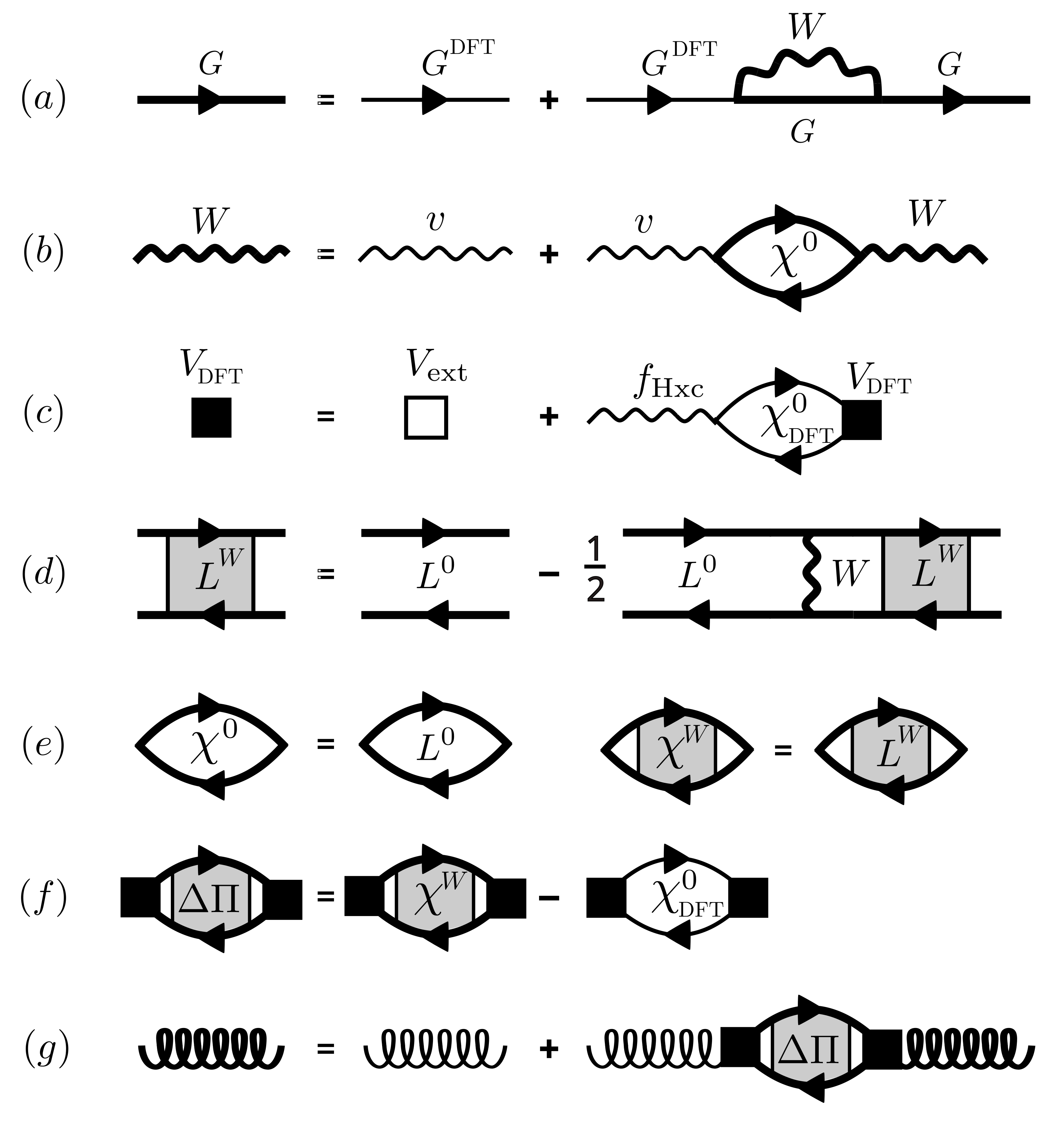}
\caption{Sketch of Feynman diagrams included in our calculations. In particular, (a) the quasi-particle Dyson equation with the nonlocal exchange term included as a self-energy, (b) the Dyson equation of the screened interaction W in the RPA, (c) the electron-phonon coupling screened by the RPA plus local exchange and correlation, (d) the Bethe-Salpeter equation with the electron-hole interaction. In (e), the definitions of $\chi^0$ and $\chi^W$ as contractions of the non-interacting $L^0$ and interacting $L^W$ electron-hole propagators.
The phonon self-energy and Dyson equation are in (f)  and (g) respectively. The Dyson equations (a) and (b) are self-consistently solved to obtain the band structure and screened interaction, while
all equations (a)-(g) are self-consistently solved to obtain the phonon frequencies and lifetimes with excitonic and dynamical effects.
}
\label{fig:feynman}
\end{figure}

%In this section, we reformulate the theoretical framework proposed in the main paper in order to exploit the Feynman diagram contributions included in the phonon calculations.
%Such Feynman diagrams are sketched in Fig. ?.
%In the present formulation, we neglect the double counting term described in Sec. , as we neglected it in our calculations. However, the extension of the proposed formulation with the double counting removal is straightforward.
%To ease the notation, we also consider a spin-unpolarized system at zero temperature. The framework can be straightforwardly generalized to spin-polarized and finite temperature.
In this Section, we reformulate the theoretical framework proposed in the main text to explicit the Feynman diagram contributions in the phonon calculation. Such Feynman diagrams are illustrated in Fig.~\ref{fig:feynman}. 
To ease notation, we neglect the double-counting term discussed in Sec.~\ref{appx:double_counting} and consider a finite system, as in the main text.
Both the double-counting term and the extension to periodic crystals are straightforward to include.

We start discussing nonlocal exchange effects for the electrons.
We define the (non-interacting) DFT electron Green's function as
\begin{equation}
    \mathcal{G}_{\DFT}(\bm r,\bm r',z) =\sum\limits_{i}\psi^{\DFT}_{i}(\bm r)\psi^{DFT*}_{i}(\bm r')\left[\frac{f^{\DFT}_{i}}{z^*-\eig^{\DFT}_{i}}
    +\frac{(1-f^{\DFT}_{i})}{z-\eig^{\DFT}_{i}} \right] \ ,
\end{equation}
where $\psi^{\DFT}_{i}$, $\eig^{\DFT}_i$ and $f^{\DFT}_{i}$ are wavefunctions, energies, and occupations obtained with DFT in a LDA or GGA functional approximation.

Nonlocal exchange effects (GDFT) can be included in on top of the DFT result solving the following Dyson equation [sketch (a) in Fig.~\ref{fig:feynman}]
\begin{equation}
    \mathcal{G}(\bm r,\bm r',z) = \mathcal{G}_{\DFT}(\bm r,\bm r\p,z)+\mathcal{G}_{\DFT}(\bm r,\bm r_1,z)\Sigma(\bm r_1,\bm r_2,z) \mathcal{G}(\bm r_2,\bm r\p,z)
\end{equation}
where $\mathcal{G}$ is the electron Green's function with nonlocal exchange effects included and the self-energy can be written as
\begin{equation}\label{eq:Sigma}
    \Sigma(\bm r,\bm r\p,z) = -n(\bm r,\bm r')W(\bm r,\bm r') \ .
\end{equation}
In the previous equation, $n$ is the electron density matrix, and $W$ is the static screened Coulomb interaction, defined below.
In the quasi-particle approximation, commonly used in the context of many-body perturbation theory~\cite{Hedin_1965,Strinati_1982,Hybertsen_1986,Godby_1988}, the interacting Green's function is 
\begin{equation}
\mathcal{G}(\bm r,\bm r',z) =\sum\limits_{i}\psi_{i}(\bm r)\psi^{*}_{i}(\bm r')
\left[\frac{f_{i}}{z^*-\eig_{i}}
+\frac{(1-f_{i})}{z-\eig_{i}} \right] \ ,
\end{equation}
where $\psi_{i}$, $\eig_i$ and $f_{i}$ are the wavefunctions, energies and occupations including nonlocal exchange effects.
In the so-called orbital approximation \cite{Reining_16} where $\psi_{i}(\bm r) \approx \psi_{i}^{\DFT}(\bm r)$, the energies are perturbatively obtained as
\begin{equation}\label{eq:eig_SX}
    \eig_{i} = \eig^{\DFT}_{i} + \mel{\psi_{i}}{\Sigma}{\psi_{i}} \ .
\end{equation}

The statically screened Coulomb interaction within the random-phase approximation (RPA) can be obtained with the following Dyson equation
\begin{equation}\label{eq:W}
    W(\bm r,\bm r\p) = v(\bm r-\bm r\p) + v(\bm r-\bm r_1)\chi^0(\bm r_1,\bm r_2,0)W(\bm r_2,\bm r\p) \ ,
\end{equation}
where $v(\bm r-\bm r\p) = e^2/|\bm r-\bm r\p|$ is the Coulomb interaction (with $e$ the electron charge) and 
\begin{equation}\label{eq:chi_0}
    \chi^0(\bm r,\bm r',z) = L^0(\bm r,\bm r,\bm r',\bm r',z) = 2\sum\limits_{ij} (f_{j}-f_{i})\frac{\psi_{i}(\bm r)\psi^*_{j}(\bm r)\psi^*_{i}(\bm r\p)\psi_{j}(\bm r\p)}{z-(\eig_{i}-\eig_{j})}
\end{equation}
is the non-interacting irreducible polarizability, i.e. the bubble diagram without interaction between the electron-hole pair.
The contraction of $L^0$ to form $\chi^0$ is sketched in the left side of the diagram (e) of Fig.~\ref{fig:feynman}.
Eqs.~\eqref{eq:Sigma}, \eqref{eq:eig_SX}, \eqref{eq:W} and \eqref{eq:chi_0} are solved self-consistently to include electron-electron interaction in $\chi^0$. This is not always a good choice, but it is appropriate for graphene~\cite{albertoTB}.
Eq.~\eqref{eq:W} is equivalent to $W(\bm r,\bm r\p) = \epsilon^{-1}(\bm r,\bm r_1)v(\bm r_1,\bm r\p)$, where $\epsilon^{-1}$ is the inverse dielectric function in the RPA.

In the partial-screen partial-screen approach~\cite{Giovanni_2024}, the electron-phonon vertices are screened by the local/semilocal DFT kernel $f^{\DFT}_{\Hxc}$ as
\begin{equation}\label{eq:V_DFT}
    V_{\DFT}^{(s\alpha)}(\bm r) = V^{(s\alpha)}_{\ext}(\bm r) + f^{\DFT}_{\Hxc}(\bm r,\bm r_1)\chi^0_{\DFT}(\bm r_1,\bm r_2)V^{(s\alpha)}_{\DFT}(\bm r_2) \ ,
\end{equation}
where $\chi^0_{\DFT}(\bm r,\bm r\p) = L^0_{\DFT}(\bm r,\bm r,\bm r\p,\bm r\p)$ is the counterpart of Eq.~\eqref{eq:chi_0} with DFT wavefunctions, energies and occupations.
Eq.~\eqref{eq:V_DFT} is represented diagramatically in  Fig.~\ref{fig:feynman}(c)  and it is equivalent to Eq.\ (10) of the main text, as $\rho_{\DFT}^{(s\alpha)}(\bm r) = \chi^0_{\DFT}(\bm r,\bm r_2)V_{\DFT}^{(s\alpha)}(\bm r_2)$.

Excitonic effects are included in the electron-hole propagator $L^W$, obtained via the Bethe-Salpeter equation (BSE)
\begin{equation}\label{eq:chiW}
L^W(\bm r,\bm r\p,\bm r\pp,\bm r\ppp,z) = L^0(\bm r,\bm r\p,\bm r\pp,\bm r\ppp,z) -\frac{1}{2}L^0(\bm r,\bm r\p,\bm r_1,\bm r_2,z)
W(\bm r_1,\bm r_2)
L^W(\bm r_1,\bm r_2,\bm r\pp,\bm r\ppp,z) \ ,
\end{equation}
where 
\begin{equation}\label{eq:L_0}
    L^0(\bm r,\bm r',\bm r'',\bm r''',z) = 2\sum\limits_{ij} (f_{j}-f_{i})\frac{\psi_{i}(\bm r)\psi^*_{j}(\bm r')\psi^*_{i}(\bm r'')\psi_{j}(\bm r''')}{z-(\eig_{i}-\eig_{j})} \ .
\end{equation}
Eq.~\eqref{eq:chiW} is represented diagramatically in Fig.~\ref{fig:feynman}(d).
The density-density response inclduing excitonic effects $\chi^W(\bm r,\bm r',z) = L(\bm r,\bm r,\bm r',\bm r',z)$ (see right side of Fig.~\ref{fig:feynman} (e)) and the partially-screened vertex $V_{\DFT}
$ are used to compute the phonon self-energy correction
\begin{equation}
    \Delta \Pi_{s\alpha,r\beta}(z) = \frac{V_{\DFT}^{(s\alpha)}(\bm r_1) \chi^W(\bm r_1,\bm r_2,z) V_{\DFT}^{(r\beta)}(\bm r_2)}{\sqrt{M_s M_r}} - \frac{V_{\DFT}^{(s\alpha)}(\bm r_1) \chi^0_{\DFT}(\bm r_1,\bm r_2,z) V_{\DFT}^{(r\beta)}(\bm r_2)}{\sqrt{M_s M_r}} \ .
\end{equation}
The Feynman diagram of the phonon self-energy is represented diagrammatically in Fig.~\ref{fig:feynman}(f).
We note the phonon self-energy has the same units  of the dynamical matrix, thus it is related to the differential force constant matrix $\Delta C$ of the main paper through the nuclear masses.

We introduce a phonon Green's function $G_{\DFT}$ without nonlocal exchange effects as
\begin{equation}
    G^{\DFT}_{s\alpha,r\beta}(z) = \sum\limits_{\l} \frac{e_{\l}^{(s\alpha)}e_{\l}^{(r,\beta)*}}{z^2-\omega^2_{\l}} \ ,
\end{equation}
where $\omega_{\l}$, $e_{\l}$ are the DFT frequencies and polarizations, satisfying the secular equation
\begin{equation}
    D^{\DFT}_{s\alpha,r\beta} e^{(r\beta)}_{\l} = \omega^2_{\l}e^{(s\alpha)}_{\l} \ ,
\end{equation}
where $D^{\DFT} = (C^{\DFT} + C^{\mathrm{ion-ion}}-\sum C^{\DFT})/\sqrt{M_sM_r}$.
The inclusion of the differential force constant term $\Delta C$ in Eq. (17) of the main paper is equivalent to solving the following Dyson equation
\begin{equation}
    G_{s\alpha,r\beta}(z) = G^{\DFT}_{s\alpha,r\beta}(z) +G_{s\alpha,p\gamma}^{\DFT}(z)\Delta\Pi_{p\gamma,t\delta}(z) G_{t\delta,r\beta}(z) \ ,
\end{equation}
[represented in Fig.~\ref{fig:feynman}(g)], in addition to the quasi-particle phonon approximation
\begin{equation} 
    G_{s\alpha,r\beta}(z) = \sum\limits_{\l} \frac{E_{\l}^{(s\alpha)}E_{\l}^{(r\beta)*}}{z^2-\Omega^2_{\l}} \ ,
\end{equation}
where $\Omega_{\l}$, $E_{\l}$ satisfy the following interacting secular equation
\begin{equation}
    [D^{\DFT}_{s\alpha,r\beta} + \Delta\Pi_{s\alpha,r\beta}(\Omega_{\l})]E_{\l}^{(r\beta)} = \Omega_{\l} E_{\l}^{(s\alpha)} \ .
\end{equation}
Finally, frequencies and inverse lifetimes can be obtained as the real and imaginary parts of $\Omega_{\l}$ respectively and the full-width half maximum is twice the inverse phonon lifetime.
This choice corresponds in taking the position of the peak and full-width half-maximum of the phonon spectral function $A(\omega)={-}\sum_{s\alpha}{\rm Im}\left[\frac{\omega}{\pi} G_{s\alpha,s\alpha}(\omega)\right] $.

In closing, we note that in Eq.\ (4) of the main text we use an alternative interaction in the form of $\mathcal{W}(\bm r,\bm r') = W(\bm r, \bm r')$ for $|\bm r-\bm r'|> r_c$, where $r_c$ a cutoff radius with the magnitude of atomic distance, and smoothly vanishing for $\bm r \to \bm r'$. This is done to  minimize the effect of the static approximation of $W$, that can be relevant for $|\bm r - \bm r'| < r_c$.

\begin{comment}
\subsection{Tight-binding model}
$\chi^0$ has been evaluated with a Dirac cone model with nonlocal effects included.
Thus, the electronic wavefunctions are given by the Dirac Hamiltonian\cite{CastroNeto_rev2009} and the band structure has the following functional form
\begin{equation}\label{eq:E_sm_model}
\begin{split}
    \eig_{\pi^*/\pi}(k) &=\pm v^{\DFT}_{\F} k \pm f v^{\DFT}_{\F} \frac{k}{2} \left[\cosh^{-1}{\left(\frac{2 k_c}{k}\right)} +\frac{1}{2}\right]  \ ,
\end{split}
\end{equation}
The values $k_c$ and $f$ are fitted over a band structure calculation sufficiently near to the Dirac point. 
The $q$ integral in Eq. () is numerically solved in elliptical coordinates to exploit the symmetry of the integrand. See Ref. [] for more details.
\end{comment}

\section{Reciprocal space for periodic systems}\label{appx:periodic}
%\subsection{Force constants: calculation method}
Here we present the expressions of the dynamical force constants [Eqs.\ (7)-(9) of the main text)] in reciprocal space. To ease of notation, we omit the double counting term in Eq. (4) of the main paper, that is straightforward to include. We consider a crystal lattice with $N$ unit cells, $N_{\text{at}}$ atoms per unit cell, and unit cell volume $\Omega$. The equilibrium position of the atom $a$ in the unit cell $I$ is $\bm R\unpert_{Is} {=} \bm R_I {+}\bm \tau_s$, $\bm{R}_I$ being a direct Bravais lattice vector and $\boldsymbol{\tau}_s$ the coordinate relative to the unit cell origin. In the main text and in Appendix I, we employed a  supercell index $s$ running from $1$ to $N_{\text{ion}}$ to indicate the nuclei. Here we pass in a crystal notation considering $N_{\text{ions}}{=}N{\times}N_{\text{at}}$ and expanding the index $s$ as $s{\to}(I,s)$.

The Fourier transform of the force constant matrix is 
\begin{equation}
    \label{C(q) = FT C(r)}
    C_{s\alpha,r\beta}(\bm q,z) = \sum_{I}e^{-i\bm q \cdot (\bm R_I+\bm \tau_s- \bm \tau_r)}  C_{s\alpha,r\beta}(\bm R_I{-}\bm R_J,z) \ ,
\end{equation}
where the sum runs on all Bravais lattice vectors and without loss of generality we have considered $\bm R_J{=}0$ \cite{Calandra_10}. In Eq.\ \eqref{C(q) = FT C(r)}, $\bm q$ is the phonon quasimomentum commensurate with a Born-Von Karman supercell of dimensions $N\Omega$.
Using \eqref{C(q) = FT C(r)} in Eq.\ 7 of the main text, we get 
\begin{equation}
    \label{C(q) = Cdft + delta C}
    C_{s\alpha,r\beta}(\bm q,z) = C^\DFT_{s\alpha,r\beta}(\bm q) + \Delta C_{r\alpha,s\beta}(\bm q,z) . 
\end{equation}
We express the response functions on the representation of the Bloch states $|\psi^{(0)}_{m\bm k}\rangle$ of quasimomentum $\bf k$, normalized on the Born–von Karman supercell of volume $N\Omega$. $|\psi^{(0)}_{m\bm k}\rangle$ are the DFT eigenfunctions, which we suppose to be the solution of the Hamiltonian with $\mathcal{W}$ included as well (also known as "orbital approximation").
The first term in Eq.\ \eqref{C(q) = Cdft + delta C} in the basis of the Bloch states reads
\begin{align} \label{C(q) DFT}
     C^\DFT_{s\alpha,r\beta}(\bm q) & = \frac{1}{N}\sum\limits_{\bm k mn} \mel{  \psi^{(0)}_{m\bm k}}{ \hat V_{\ext}^{(s\alpha,-\bm q)}}{\psi^{(0)}_{n\bm k +\bm q}}  \mel{\psi^{(0)}_{n\bm k +\bm q}}{\hat \rho_{\DFT}^{(r\beta,\bm q)}}{\psi^{(0)}_{m\bm k}},
\end{align}
where a caret over the symbol indicates quantum mechanical operators. 
The induced external potential $\hat V_{\ext}^{(s\alpha,-\bm q)}$ in Eq.\ \eqref{C(q) DFT} is defined as\cite{Giovanni_2024}
\begin{align} \label{Vext (q)}
    \hat V_{\ext}^{(s\alpha,-\bm q)}&=\sum_{\bm{R}}e^{-i\bm q \cdot \left(\bm R+\boldsymbol{\tau}_s\right)}\frac{\partial \hat V_{\text{ext}}}{\partial u{(\bm{R})}_{s\alpha}}{=}\hat U_\ext^{(s\alpha,-\bm q)}e^{-i\bm q\cdot \hat{\bm {r}}} ,         
\end{align}
where $\hat V_\ext$ was defined in Schrodinger representation in Eq.\ \eqref{v ext app}, $\hat U_\ext^{(r\beta,\bm q)}$ are cell-periodic potentials, and the sum over $\bm R$ runs over the full crystal.

The DFT induced density in \eqref{C(q) DFT} is obtained as the solution of a static self-consistent set of Sternheimer equations 
\begin{align}\label{app: sternheimer}
\mel{\psi^{(0)}_{n\bm k +\bm q}}{\hat \rho_{\DFT}^{(r\beta,\bm q)}}{\psi^{(0)}_{m\bm k}}&=
\left[\chi^{0}_\DFT{}\right]_{n\bm k {+} \bm q, m\bm k}(0)
  \mel{\psi^{(0)}_{n\bm k +\bm q}}{ \hat V_{\DFT}^{(r\beta,\bm q)}}{\psi^{(0)}_{m\bm k}} ,\\
 \mel{\psi^{(0)}_{n\bm k +\bm q}}{ \hat V_{\DFT}^{(r\beta,\bm q)}}{\psi^{(0)}_{m\bm k}}&= 
\mel{ \psi^{(0)}_{n\bm k +\bm q}}{ \hat V_{\ext}^{(r\beta,\bm q)}}{\psi^{(0)}_{m\bm k}}
  {+}\frac{1}{N}\sum_{\bm k\p sl}
 \left[f_{\Hxc}\right]^{n\bm k {+} \bm q, m\bm k}_{p\bm k' {+} \bm q, l\bm k'}
\mel{\psi^{(0)}_{p\bm k' + \bm q}}{ \hat \rho_{\DFT}^{(r\beta,\bm q)}}{\psi^{(0)}_{l\bm k'}}.
 \label{app: V_scf = V + Kn}
\end{align}
In Eq.\ \eqref{app: sternheimer}, the frequency dependent noninteracting Kohn-Sham polarizability is defined as
\begin{align}
\left[\chi^{0}_\DFT{}\right]_{n\bm k {+} \bm q, m\bm k}(z)
   &= 2 \frac{f_{m\bm k}^\DFT - f_{n\bm k +\bm q}^\DFT}{ \hbar z -(\varepsilon^\DFT_{n\bm k+\bm q}-\varepsilon^\DFT_{m\bm k})}, \label{chi 0 DFT def}
\end{align}
where $f^\DFT_{m\bm k}, \varepsilon^\DFT_{m\bm k}$ are respectively the Fermi-Dirac occupation and the energy of the Kohn Sham state of band $m$ and quasimomentum $\bm k$.
In Eq.\ \eqref{app: V_scf = V + Kn} the Hartree-exchange correlation kernel in reciprocal space is defined as
\begin{equation}
    \label{f Hxc (q)}
  \left[f^{\DFT}_{\Hxc}\right]^{n\bm k {+} \bm q, m\bm k}_{p\bm k' {+} \bm q, l\bm k'}=   N\!\!\int\limits_{N\Omega}\!\! \dd{\bm r} 
\psi^{(0)}_{n\bm k+\bm q}(\bm r)^*\psi^{(0)}_{m\bm k}(\bm r)f^{\DFT}_{\Hxc}(\bm r,\bm r_1)\psi^{(0)}_{p\bm k'+\bm q}(\bm r_1) \psi^{(0)}_{l\bm k\p}(\bm r_1)^*
\end{equation}
where the expression of the kernel in real space $f^{\DFT}_{\Hxc}(\bm r,\bm r')$ depends on the approximation used for the Hartree-exchange correlation energy functional. 

The second term in Eq.\ \eqref{C(q) = Cdft + delta C} is expressed in reciprocal space as

\begin{equation} \label{DeltaC (q)}
\Delta C_{s\alpha, r\beta}(\bm q,z)= \frac{1}{N}\sum_{\bm k n  m }
\mel{  \psi^{(0)}_{m\bm k}}{ \hat V^{(s\alpha,-\bm q)}_\DFT}{\psi^{(0)}_{n\bm k +\bm q}}
\left[ \chi^{W}_{n\bm k {+} \bm q, m\bm k}(z) - \left[\chi^{0}_\DFT{}\right]_{n\bm k {+} \bm q, m\bm k}(z)\right]
\mel{  \psi^{(0)}_{n\bm k +\bm q}} { \hat V^{(r\beta,\bm q)}_\DFT}{\psi^{(0)}_{m\bm k}}.
\end{equation}
The second term in Eq.~\eqref{DeltaC (q)} is trivial to compute. In the remainder, we focus in the calculation of the first term.

The interacting electron-hole propagator $L^W$, required to obtain its contracted counterpart $\chi^W$,  satisfies the following BSE in reciprocal space
\begin{equation}\label{eq:BSE_Tspace}
    [L^{W}]^{n\bm k {+} \bm q, m\bm k}_{p\bm k' {+} \bm q, l\bm k\p}(z) = 
    L^{0}_{p\bm k' {+} \bm q, l\bm k\p}(z)\delta_{\k nm,\k\p pl} +
    \frac{1}{N}\sum\limits_{\k\pp vt}
    L^{0}_{p\bm k' {+} \bm q, l\bm k\p}(z)
    [W]^{p\bm k' {+} \bm q, l\bm k\p}_{v\bm k\pp {+} \bm q, t\bm k\pp}
    [L^{W}]_{v\bm k\pp {+} \bm q, t\bm k\pp}^{n\bm k {+} \bm q, m\bm k}(z),
\end{equation}
where
\begin{equation}
    L^{0}_{n\bm k {+} \bm q, m\bm k}(z)
   = 2 \frac{f_{m\bm k} - f_{n\bm k +\bm q}}{ \hbar z -(\varepsilon_{n\bm k+\bm q}-\varepsilon_{m\bm k})} \label{L 0 def}
\end{equation}
and
\begin{equation}
    [W]^{n\bm k {+} \bm q, m\bm k}_{p\bm k' {+} \bm q, l\bm k\p} = \!\!\int\limits_{N\Omega}\!\! \dd{\bm r}\psi^{(0)}_{n\k+\qph}(\bm r_1)^*\psi^{(0)}_{m\k}(\bm r_2) W(\bm r_1,\bm r_2)\psi^{(0)}_{s\k\p+\qph}(\bm r_1)\psi^{(0)}_{l\k\p}(\bm r_2)^* \ .
\end{equation}
By multiplying Eq.~\eqref{eq:BSE_Tspace} on the right side by $\mel{\psi^{(0)}_{s\k\p+\qph}}{V_{\DFT}^{(r\beta,\qph)}}{\psi^{(0)}_{l\k\p}}$ and summing over $\k\p sl$ we find
\begin{equation}\label{eq:Stern_W}
    \mel{\psi^{(0)}_{n\bm k +\bm q}}{\hat \rho^{(r\beta,\bm q)}(z)}{\psi^{(0)}_{m\bm k}} =
    \mel{\psi^{(0)}_{n\bm k +\bm q}}{\hat \rho_0^{(r\beta,\bm q)}(z)}{\psi^{(0)}_{m\bm k}}+
    \frac{1}{N}\sum\limits_{\k\pp vt}
    L^{0}_{n\bm k {+} \bm q, m\bm k}(z)
    [W]^{n\bm k {+} \bm q, m\bm k}_{v\bm k\pp {+} \bm q, t\bm k\pp}
    \mel{\psi^{(0)}_{v\bm k\pp +\bm q}}{\hat \rho^{(r\beta,\bm q)}(z)}{\psi^{(0)}_{t\bm k\pp}},
\end{equation}
where $\rho = L^{W} V_{\DFT}$ and $\rho_0=L^0V_{\DFT}$ are respectively the electron densities with and without excitonic effects.
The first term of the differential force constants in Eq.~\eqref{DeltaC (q)} can be written in terms of the density with excitonic effects as
\begin{equation}\label{eq:C_1}
\sum_{\bm k n  m }
\mel{  \psi^{(0)}_{m\bm k}}{ \hat V^{(s\alpha,-\bm q)}_\DFT}{\psi^{(0)}_{n\bm k +\bm q}}
 \chi^{W}_{n\bm k {+} \bm q, m\bm k}(z)
\mel{  \psi^{(0)}_{n\bm k +\bm q}} { \hat V^{(r\beta,\bm q)}_\DFT}{\psi^{(0)}_{m\bm k}}= \sum_{\bm k n  m }
\mel{  \psi^{(0)}_{m\bm k}}{ \hat V^{(s\alpha,-\bm q)}_\DFT}{\psi^{(0)}_{n\bm k +\bm q}}\mel{\psi^{(0)}_{n\bm k +\bm q}}{\hat \rho^{(r\beta,\bm q)}(z)}{\psi^{(0)}_{m\bm k}}.
\end{equation}
We solve Eq.~\eqref{eq:Stern_W} iteratively using $\rho\pert_{\text{start}}= \chi\unpert V_\ext\pert$ as a starting guess. We iterate until convergence of Eq.~\eqref{eq:C_1}. Between different iterations, we use a  linear mixing procedure to update the density matrix \cite{albertoTB,Giovanni_2024}.

%%% da finire

Using Eq.\ \eqref{C(q) DFT} and Eq.\ \eqref{DeltaC (q)} in Eq.\ \eqref{C(q) = Cdft + delta C}, we obtain $C_{s\alpha,r\beta}(\bm q,z)$ as a $N_dN_{\text{at}} {\times}N_dN_{\text{at}}$ symmetric matrix for every momentum $\bm q$ and frequency $z$, $N_d$ being the dimensionality of the crystal (in our 2D model of  graphene \cite{albertoTB}, $C_{s\alpha,r\beta}(\bm q,z)$ is a $4{\times}4$ matrix for every $(\bm q, z)$). The phonon dispersion $\Omega(\bm q)_\lambda$ is obtained as the solution of the secular equation 
$\sum_{r\beta}C_{s\alpha,r\beta}(\bm q,\Omega(\bm q)_\lambda)e(\bm q)_{\lambda,r\beta} = \Omega(\bm q)_\lambda e(\bm q)_{\lambda,s\alpha}$.

\begin{comment}
%%================================%%
\subsection{Fock-like double counting exchange term}
Finally, we provide an expression of the double-counting exchange energy density $\varepsilon_{\xc}^{\mathrm{dc}}$ in Eq.~\eqref{eq:e_x^dc} in Fourier space.
The density matrix of the uniform electron gas with Fermi wavevector $k_{\F} = (6\pi^2\rho)^{1/3}$ can be written as
\begin{equation}\label{eq:n_HEG_def}
    n^{\mathrm{HEG}}[\rho](\bm r,\bm r\p) = \sum\limits_{|\k|<k_{\F}} \frac{e^{i\k\cdot(\bm r-\bm r\p)}}{V} \ ,
\end{equation} 
where $V$ is the total integration volume. The translationally invariant screened interaction $\mathcal{W}(\bm r-\bm r\p)$ can be written in term of its Fourier transform as
\begin{equation}\label{eq:W_q}
    \mathcal{W}(\bm r-\bm r\p) = \frac{1}{\Omega}\sum\limits_{\qph\G}
    \mathcal{W}(\qph+\G)e^{i(\qph+\G)\cdot(\bm r-\bm r\p)} \ .
\end{equation}
By substituting Eqs.~\eqref{eq:n_HEG_def} and \eqref{eq:W_q} into Eq.~\eqref{eq:e_x^dc}, we find
\begin{equation}
    \varepsilon^{\mathrm{dc}}_{\x}[\rho] = \frac{\rho}{2V^2}
    \sum\limits_{|\k|, |\k\p|< k_{\F}} \mathcal{W}(\k-\k\p) \ .
\end{equation}
The double counting energy density can be obtained for each density $\rho$, then used to evaluate $E_{\Hxc}^{\mathrm{dc}}$.
\end{comment}

\section{Gauge field at $\mathbf{q}=0$}
In a low-energy Dirac cone model, the static interaction between electrons and phonons at $\mathbf{q}=0$ can be described via a gauge-field coupling \cite{PhysRevB.76.045430, Bistoni_2019, PhysRevB.110.L201405, PhysRevB.110.L201405}. We restrict our analysis around the K point, even though similar arguments hold for K' point as well. In particular, near the K point the Hamiltonian can be written in the form:
\begin{align}
\boldsymbol{\mathcal{A}}=\sum_s \boldsymbol{\mathcal{A}}_s \quad \boldsymbol{\mathcal{A}}_s=l_s \hat{\mathbf{z}} \times \mathbf{u}_s \quad l_s=\pm 1,\\
H_{\mathbf{k}}=\hbar v_{\textrm{F}}\left(k_x+\frac{\beta}{b_0^2}A_x,k_y+\frac{\beta}{b_0^2}A_y,0\right) \cdot \boldsymbol{\sigma}^{\textrm{P}} \label{eq:Hamilgauge}
\end{align}
where the sum over $s$ is the sum over the atoms of graphene, $ \mathbf{u}_s$ are atomic displacements with the same periodicity of the lattice (i.e. corresponding to $\mathbf{q}=0$), $b_0=a_0/\sqrt{3}$ is the interatomic distance ($a_o$ being the lattice parameter), and $\boldsymbol{\sigma}^{\textrm{P}}$ is the vector containing the pseudospin Pauli matrices. $\beta$ is an adimensional coupling constant. From the model of Eq. \ref{eq:Hamilgauge}, it follows that the phonon distortion causes a lateral displacement of the Dirac cone, and therefore it induces a gap opening which is proportional to the Fermi velocity. More precisely
\begin{align}
\frac{\partial \epsilon_{\pi/\pi^* \mathbf{K}}}{\partial u_{sx}}=\langle u_{\pi/\pi^* \mathbf{K}}| \frac{\partial H_{\mathbf{k}}}{\partial u_{sx}}| u_{\pi/\pi^* \mathbf{K}} \rangle =l_s \frac{\beta}{b_0^2} \langle u_{\pi/\pi^* \mathbf{K}}| \frac{\partial H_{\mathbf{k}}}{\partial k_{y}}| u_{\pi/\pi^* \mathbf{K}} \rangle = \pm  l_s \frac{\beta}{b_0^2} v_{\textrm{F}}, \label{eq:energgauge}
\end{align}
and similarly for the derivative with respect to $u_{sy}$. It is evident that the electron-phonon matrix element is here proportional to the Fermi velocity via the adimensional coupling $\beta$. 

The Hamiltonian description of Eq.\ \eqref{eq:Hamilgauge} is shown for the case of constant Fermi velocity. Though, under proper hypothesis, it can be extended to the case where the band velocity is renormalized by the electron-electron interaction. In this case, the full Hamiltonian reads
\begin{align}
H_{\mathbf{k}}=H^0_{\mathbf{k}}+H^{\textrm{e-ph}}_{\mathbf{k}}+H^{\textrm{e-e}}_{\mathbf{k}},\, H_0=\hbar v_{\textrm{F}}\left(k_x,k_y,0\right) \cdot \boldsymbol{\sigma}^{\textrm{P}}, \, H^{\textrm{e-ph}}_{\mathbf{k}}=\hbar v_{\textrm{F}}\left(\frac{\beta}{b_0^2}A_x,\frac{\beta}{b_0^2}A_y,0\right) \cdot \boldsymbol{\sigma}^{\textrm{P}}.
\end{align}
To write the above expression, we implicitly assumed that the electron-electron interaction is static. Then, we can again determine the strength of the electron-phonon interaction, now dressed by the electron-electron interaction, by evaluating the change in band energies with respect to a displacement. In the approximation of cylindrical isotropy of our low energy model, the action of $H^{\textrm{e-ph}}_{\mathbf{k}}$ still consists in shifting Dirac cone center origin from $\mathbf{K}$ to $\mathbf{K(u)}$. If we assume that the action of $H^{\textrm{e-e}}_{\mathbf{k}}$ is independent of the origin of the Dirac cone, i.e. that $H^{\textrm{e-e}}_{\mathbf{k}}=H^{\textrm{e-e}}(|\mathbf{k}-\mathbf{K(u)}|)$, then the action of the dressed electron-phonon is still equivalent to a gauge coupling. In other words, the action of the dressed electron-phonon coupling is still to shift the modified band dispersion from $\mathbf{K}$ to $\mathbf{K(u)}$. Of course, the above assumptions are not exact for realistic models, but are expected to be satisfied to a satisfactory extent for very small $\mathbf{u}$. We therefore expect
\begin{align}
\frac{\partial \epsilon_{\pi/\pi^* \mathbf{K+k}}}{\partial u_{sx}} \sim \pm  l_s \frac{\beta}{b_0^2} v_{\textrm{g}}(\mathbf{k}),\label{eq:protection}
\end{align}
These considerations practically provides a `gauge field protection' for the linewidth value, as discussed in the next section.
\section{Phonon lifetimes at $\Gamma$ and K within logarithmic Dirac cone model}
\label{sec:ladderlinewidth}
In this section, we derive an expression of the phonon lifetimes at $\Gamma$ and K with the assumption that the electronic bands are isotropic around K, thus
\begin{equation}
    \varepsilon_{\pi^*/\pi\k} =  \pm  v_{\phi}(\kappa) \hbar\kappa
\end{equation}
where $\kappa = |\k - \mathrm{K}|$ and $v_{\phi}(\kappa)$ is the electron phase velocity, $\kappa$ dependent to include the logarithmic divergence in the case of exact-exchange corrections~\cite{albertoTB}.

The general expression of the  phonon full-width half maximum (FWHM) in a 2D crystal is~\cite{Giovanni_2024}
\begin{equation}
    \Gamma_{\l\qph}(\omega_{\l\qph}) = \frac{2\pi}{\hbar}2\frac{1}{N}\sum\limits_{\k nm}(f_{m\k}-f_{n\k+\qph})\left|
    g_{nm}^{\l\qph}(\k)\right|^2 \delta[\hbar\omega_{\l\qph}-(\varepsilon_{n\k+\qph}-\varepsilon_{m\k})] \ ,
\end{equation}
where
\begin{equation}
  g_{nm}^{\l\qph}(\k) = \sum\limits_{s\alpha}\mel{\psi_{n\k+\qph}^{(0)}}{V_{\mathrm{scf}}^{(s\alpha)}}{\psi^{(0)}_{m\k}}\frac{e^{(s\alpha)}_{\lambda\mathbf{q}}}{\sqrt{2M_{\mathrm{c}}\omega_{\l\qph}/\hbar}} \ ,
\end{equation}
where $V_{\mathrm{scf}}$ is the self-consistent potential induced by the phonon perturbation (which can be both from DFT or GDFT).
We focus on the TO mode at K and the LO mode at $\Gamma$. We omit the branch index to ease the notation.
In the range $\hbar\omega_{\indxlt}\gg K_{\mathrm{B}}T$, only $n=\pi^*$, $m=\pi$ contributions are relevant.
We also consider the continuous limit of the $\k$ summation and change the variable of integration to $\kappa$ and find
\begin{equation}\label{eq:Gamma_init}
    \Gamma_{\indxlt}(\omega_{\indxlt}) =
    \frac{2\pi}{\hbar}2Ad_{\indxlt} \int \frac{d^2\kappa}{(2\pi)^2} |g^{\indxlt}(\bm \kappa)|^2
    \delta[\hbar\omega_{\indxlt}-2\hbar v_{\phi}(\kappa)\kappa] \ ,
\end{equation}
where $g^{\indxlt}(\bm \kappa) = {g_{\pi^* \pi}^{\textrm{LO},\Gamma / \textrm{TO,K}}(\k)}$, $d_{\indxlt} = 2/1$ at $\indxlt$ is the Dirac cone degeneracy and $A$ is the area of the graphene unit cell.
The integral in Eq.~\eqref{eq:Gamma_init} can be written in polar coordinates as
\begin{equation}
    \Gamma_{\indxlt}(\omega_{\indxlt}) =
    \frac{2\pi}{\hbar}2Ad_{\indxlt} \frac{2\pi}{(2\pi)^2}  \int d\kappa \ \kappa |\bar{g}^{\indxlt}( \kappa)|^2
    \delta[\hbar\omega_{\indxlt}-2\hbar v_{\phi}(\kappa)\kappa] \ ,
\end{equation}
where
\begin{equation}
    |\bar{g}^{\indxlt}(\kappa)|^2 = \int \frac{d\theta}{2\pi} |g^{\indxlt}(\bm \kappa)|^2
\end{equation}
is the electron-phonon coupling matrix element angle-averaged at a fixed distance from the Dirac point.
We now make another change of coordinates: $E = 2\hbar v_{\phi}(\kappa)\kappa$, $dE = 2\hbar v_{\mathrm{g}}(\kappa)d\kappa$, where $\hbar v_{\mathrm{g}}(\kappa) = d\varepsilon_{m\kappa}/d\kappa = \hbar v_{\phi}(\kappa)+ dv_{\phi}(\kappa)/d\kappa \ \hbar \kappa$ is the radial electron group velocity.
Within the proposed change of coordinates, the $\Gamma$ can be written as
\begin{equation}
    \Gamma_{\indxlt}(\omega_{\indxlt}) =
    \frac{2\pi}{\hbar}2Ad_{\indxlt} \frac{2\pi}{(2\pi)^2}  \int \frac{dE}{2\hbar v_{\mathrm{g}}(\kappa)} \ \frac{E}{2\hbar v_{\phi}(\kappa)} |\bar{g}^{\indxlt}( \kappa)|^2\delta(\hbar\omega_{\indxlt}-E).
\end{equation}
By evaluating the Dirac delta, we obtain
\begin{equation}
    \Gamma_{\indxlt}(\omega_{\indxlt}) =
    \frac{Ad_{\indxlt}}{2\hbar}   \frac{|\bar{g}^{\indxlt}( \bar{\kappa}_{\indxlt})|^2\hbar\omega_{\indxlt}}{\hbar v_{\mathrm{g}}(\bar{\kappa}_{\indxlt})\hbar v_{\phi}(\bar{\kappa}_{\indxlt})} \ ,
    \label{eq:Gammamodel}
\end{equation}
where $\bar{\kappa}$ is the wave-vector distance from the Dirac point where $\hbar\omega_{\indxlt} = 2\varepsilon_{\pi\bar{\kappa}_{\indxlt}} = 2\varepsilon_{\pi^*\bar{\kappa}_{\indxlt}}$
As the joint density of states is
\begin{equation}
    J_{\indxlt}(\hbar\omega_{\indxlt}) = d_{\indxlt} A \frac{1}{2\pi} \frac{\hbar\omega_{\indxlt}}{2\hbar v_{\mathrm{g}}(\bar{\kappa}_{\indxlt})\hbar v_{\phi}(\bar{\kappa}_{\indxlt})},
\end{equation}
the $\Gamma$ can be written as
\begin{equation}
    \Gamma_{\indxlt}(\omega_{\indxlt}) =
    \frac{2\pi}{\hbar}|\bar{g}^{\indxlt}( \bar{\kappa}_{\indxlt})|^2
    J_{\indxlt}(\hbar \omega_{\indxlt}) \ .
    \label{eq:linescaling}
\end{equation}
\begin{comment}
The full-width half maximum (FWHM) is expressed
\begin{equation}
    \gamma_{\indxlt} = \frac{\Gamma_{\indxlt}(\omega_{\indxlt})}{2M_{\mathrm{C}}\omega_{\indxlt}}
\end{equation}

\begin{equation}
    \gamma_{\indxlt} = g_{\indxlt} \frac{A}{8M_{\mathrm{C}}}
    \frac{|\bar{V}_{\mathrm{scf}}^{\indxlt}( \bar{\kappa})|^2}{v_{\mathrm{G}}(\bar{\kappa})v_{\mathrm{\phi}}(\bar{\kappa})}
\end{equation}
$v_{\G}$
\end{comment}

\begin{figure}[h!]
\includegraphics[width=\linewidth/4*3]{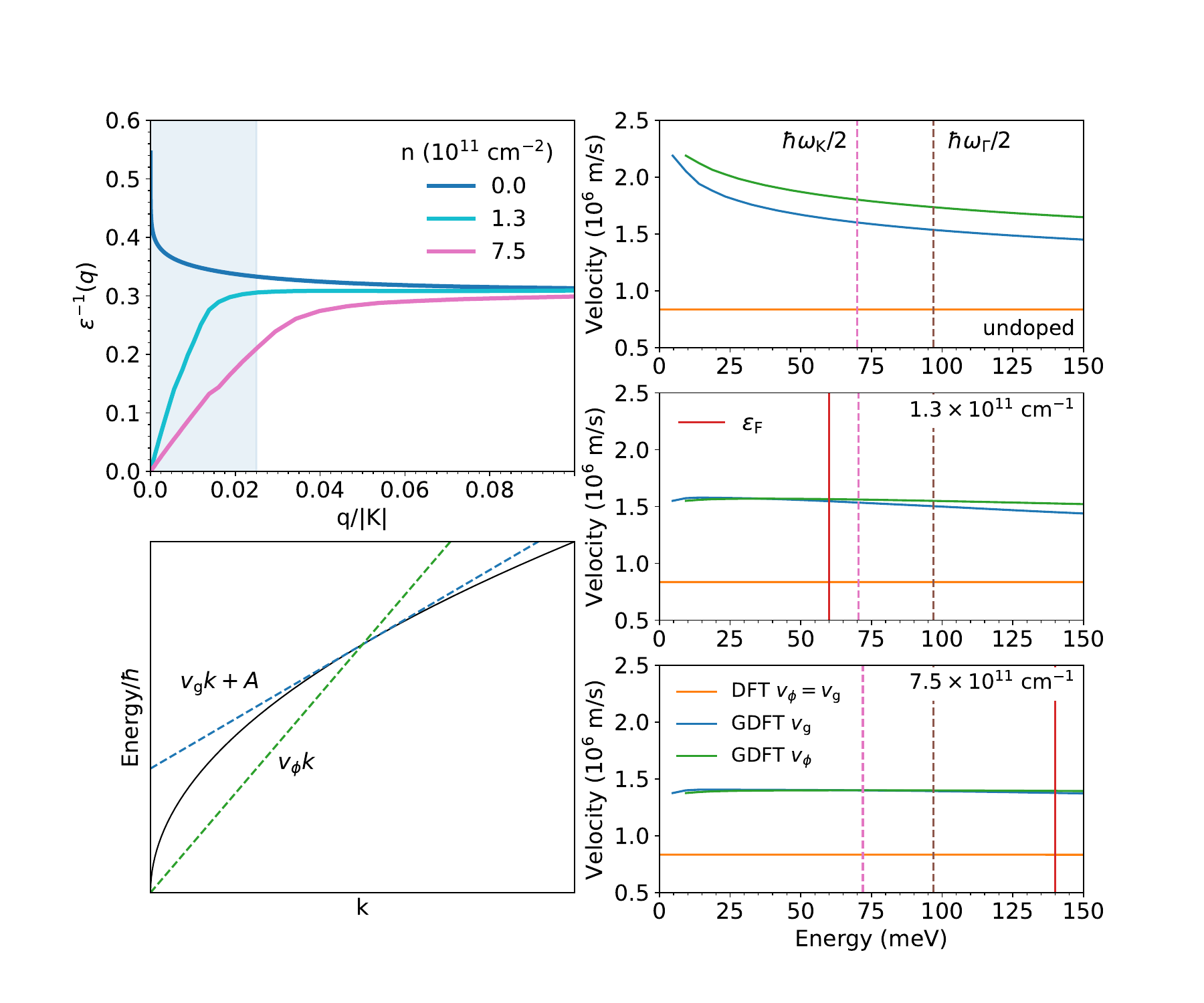}
\caption{Top left: static inverse dielectric functions in the random phase approximation used to calculate the static screened interaction $\mathcal{W}$ for undoped and the doping levels used in the main text. The blue-colored region indicated the momentum regime where dynamical corrections in optical phonons are relevant.
Bottom left: sketch of the definition of the group and phase velocities. For a complex band structure, like the one of graphene, we always have $v_{\phi} > v_{\mathrm{g}}$. Right: group ($v_{\mathrm{g}}$) and phase ($v_{\phi}$) velocities of electrons as a function of the band energy for undoped and the doping levels studied in the main text.}
\label{fig:velocities}
\end{figure}

In Tab. \ref{Tab_CC} we present the results for $\Gamma_{\Gamma,\mathrm{K}}$ computed via the methodology explained in the main text. We also report $\bar{\kappa}_{\indxlt}$, $v_{g}(\bar{\kappa}_{\indxlt})$ and $v_{\phi}(\bar{\kappa}_{\indxlt})$ computed at different levels of theory; the modulus of the velocities are obtained as an average between their value along $\mathrm{K}-\Gamma$ and $\mathrm{K}-\mathrm{M}$, and between $\pi^*$ and $\pi$ bands (see Fig. \ref{fig:velocities}). Then, we finally use Eq. \ref{eq:Gammamodel} to deduce $|\bar{g}_{\mathrm{scf}}^{\indxlt}( \bar{\kappa}_{\indxlt})|^2$. The following relations hold with the definition of Ref. \onlinecite{Piscanec_2004}
\begin{align}
|\bar{g}^{\Gamma}( \bar{\kappa}_{\Gamma}\rightarrow 0)|^2=  \langle g^2_{\Gamma} \rangle_{\mathrm{F}},\quad |\bar{g}^{\mathrm{K}}( \bar{\kappa}_{\mathrm{K}}\rightarrow 0)|^2=2 \langle g^2_{\mathrm{K}} \rangle_{\mathrm{F}},
\end{align}
where a factor of $2$  at zone border comes from the angular dependence of the vertex.

We also report the values for the deformation potential defined as
\begin{align}
|\bar{\mathcal{D}}^{\indxlt}( \bar{\kappa}_{\indxlt})|^2=\frac{2}{\hbar}M_c\omega_{\indxlt}|\bar{g}^{\indxlt}( \bar{\kappa}_{\indxlt})|^2,
\end{align}
and of the adimensional parameter (accordingly to Eq. \ref{eq:protection}) 
\begin{align}
\beta=\frac{b_0^2}{\hbar v_{g}(\bar{\kappa}_{\Gamma})}|\bar{\mathcal{D}}^{\Gamma}(\bar{\kappa}_{\Gamma})|.
\end{align}

By the values of Tab. \ref{Tab_CC} and using the same notation for bands and linear response, we find, for zero doping 
\begin{align}
&\left(\frac{|\bar{\mathcal{D}}^{\mathrm{K}}( \bar{\kappa}_{\mathrm{K})}|^2 }{4|\bar{\mathcal{D}}^{\Gamma}( \bar{\kappa}_{\Gamma})|^2 }\right)_{\textrm{DFT}}\sim 1, \quad 
\left(\frac{|\bar{\mathcal{D}}^{\mathrm{K}}( \bar{\kappa}_{\mathrm{K})}|^2 }{4|\bar{\mathcal{D}}^{\Gamma}( \bar{\kappa}_{\Gamma})|^2 }\right)_{\mathrm{GDFT}}\sim 4.6, \\
&\left(\frac{\beta^{\textrm{K}}(\bar{\kappa}_{\textrm{K}})}{\beta^{\Gamma}(\bar{\kappa}_{\Gamma})}\right)_{\textrm{DFT}}\sim 2, \quad \left(\frac{\beta^{\textrm{K}}(\bar{\kappa}_{\textrm{K}})}{\beta^{\Gamma}(\bar{\kappa}_{\Gamma})}\right)_{\mathrm{GDFT}} \sim 4,
\end{align}
which shows a large enhancement of the coupling at the K point with respect to $\Gamma$ ascribable to the effect of electronic vertex corrections. For the doping level shown in table, the enhancements are still present even if slightly reduced.

Notice that the invariance of the linewidths at $\mathbf{q}$=0 with respect to the response theory level is ascribable to an almost exact compensation between electron-phonon and velocities scalings in Eq. \ref{eq:linescaling}, as evident from the results of Tab. \ref{Tab_CC}. This behaviour is inherited from the `gauge-field protection' of the Dirac cone model due to Eq. \ref{eq:energgauge} discussed in the previous Section.

\begin{table}
\centering
\begin{ruledtabular}
\begin{tabular}{ccccccccc||ccccc}
%\\[-3pt]
\multicolumn{14}{c}{graphene undoped}\\
\\[-8pt]
Theory  & $\hbar \Gamma_{\Gamma}$ & $\hbar \Gamma_{\mathrm{K}}$ & $\bar{\kappa}_{\Gamma}$ & $\bar{\kappa}_{\mathrm{K}}$  & $v_{g}(\bar{\kappa}_{\Gamma})$   & $v_{g}(\bar{\kappa}_{\mathrm{K}})$  & $v_{\phi}(\bar{\kappa}_{\Gamma})$  & $v_{\phi}(\bar{\kappa}_{\mathrm{K}})$ & $|\bar{g}^{\mathrm{\Gamma}}(\bar{\kappa}_{\Gamma})|^2$  & $|\bar{g}^{\mathrm{K}}(\bar{\kappa}_{\mathrm{K}})|^2$ & $|\bar{\mathcal{D}}^{\mathrm{\Gamma}}(\bar{\kappa}_{\Gamma})|^2$  & $|\bar{\mathcal{D}}^{\mathrm{K}}(\bar{\kappa}_{\mathrm{K}})|^2$ & $\beta(\bar{\kappa}_{\Gamma})$ \\
   & \multicolumn{2}{c}{(cm$^{-1}$)} & \multicolumn{2}{c}{($10^{-3}$ \AA$^{-1}$)} & \multicolumn{4}{c}{($10^6$ m/s)}  &\multicolumn{2}{c}{(eV$^2$)}& \multicolumn{2}{c}{(eV$^2/$\AA$^2$)} & \multicolumn{1}{c}{Adim.}  \\
\hline
DFT               & $10.86$ & $21.75$ & $17.63$ & $14.43$ & $0.83$ & $0.83$ & $0.82$ & $0.83$ & $0.039$ & $0.193$ & $44.22$ & $177.75$& $2.50$ \\
%GW$_{\mathrm{st}}$& NL & $ 2.32$ & $ 3.63$ & $  7.71$ & $ 5.36$ & $1.70$ & $1.78$ & $1.90$ & $1.97$ & $0.039$ & $0.194$ & $44.27$ & $150.62$\\
GDFT& $11.53$ & $97.34$ & $ 7.73$ & $ 5.38$ & $1.70$ & $1.78$ & $1.90$ & $1.97$ & $0.197$ & $4.992$ & $220.37$ & $4041.11$ & $2.67$ \\
%\\[-3pt]
\hline
\hline
\\[-3pt]
\hline
\hline
\\[-8pt]
\multicolumn{14}{c}{graphene doped ($n = 1.3\times 10^{11}$ cm$^{-1}$, $[E_{\textrm{F}}]_{\textrm{DFT}}=0.035$eV, $[E_{\textrm{F}}]_{\mathrm{GDFT}}=0.066$eV)}\\
\\[-8pt]
%\\[-3pt]
Theory  & $\hbar \Gamma_{\Gamma}$ & $\hbar \Gamma_{\mathrm{K}}$ & $\bar{\kappa}_{\Gamma}$ & $\bar{\kappa}_{\mathrm{K}}$  & $v_{g}(\bar{\kappa}_{\Gamma})$   & $v_{g}(\bar{\kappa}_{\mathrm{K}})$  & $v_{\phi}(\bar{\kappa}_{\Gamma})$  & $v_{\phi}(\bar{\kappa}_{\mathrm{K}})$ & $|\bar{g}^{\mathrm{\Gamma}}(\bar{\kappa}_{\Gamma})|^2$  & $|\bar{g}^{\mathrm{K}}(\bar{\kappa}_{\mathrm{K}})|^2$ & $|\bar{\mathcal{D}}^{\mathrm{\Gamma}}(\bar{\kappa}_{\Gamma})|^2$  & $|\bar{\mathcal{D}}^{\mathrm{K}}(\bar{\kappa}_{\mathrm{K}})|^2$ & $\beta(\bar{\kappa}_{\Gamma})$ \\
   & \multicolumn{2}{c}{(cm$^{-1}$)} & \multicolumn{2}{c}{($10^{-3}$ \AA$^{-1}$)} & \multicolumn{4}{c}{($10^6$ m/s)}  &\multicolumn{2}{c}{(eV$^2$)}& \multicolumn{2}{c}{(eV$^2/$\AA$^2$)} & \multicolumn{1}{c}{Adim.} \\
\hline
DFT              & $10.84$ & $21.63$ & $14.43$ & $12.81$ & $0.83$ & $0.83$ & $0.82$ & $0.83$ & $0.039$ & $0.192$ & $44.14$ & $176.60$& $2.50$\\
GDFT & $11.85$ & $81.83$ & $ 9.39$ & $ 6.82$ & $1.56$ & $1.57$ & $1.56$ & $1.56$ & $0.152$ & $2.912$ & $170.05$ & $2373.01$& $2.59$\\
\end{tabular}
\end{ruledtabular}
\caption{Numerical values of quantities evaluated at $q=\Gamma,\mathrm{K}$ for different levels of theory implemented in the calculation of band energies and linear response.}
\label{Tab_CC}
\end{table}

\section{Computational details---extended}

%\textcolor{red}{[aggiungere questione avoided crossing]}
The DFT band structure and force constants $C^{\mathrm{ion-ion}}$ and $C^{\mathrm{KS}}$ are obtained with the QUANTUM ESPRESSO package~\citep{QE_2020} within Perdew-Burke-Erzenov (PBE) approximation \cite{Perdew_1996}.
We adopt norm-conserving pseudopotentials to model the electron-ion interaction and a kinetic energy cutoff of $90$ Ry.
The force constants are calculated with DFPT using a Sternheimer approach. The screened interaction $W$ is modeled with Dirac-cone screening in the random-phase approximation and calculated self-consistently with the band structure~\cite{albertoTB}

The excitonic and/or dynamical corrections $\Delta C(z)$ are calculated within a tight-binding approach using
an in-house built python3 code. We employ the five nearest-neighbor tight-binding model of Ref.~\onlinecite{albertoTB}, including only $\pi$/$\pi^*$ bands, with the parameters of the model fitted over the ab-initio band structure. We use the fitted model to compute $\chi^0_{\DFT}$ on fine grids. In addition, we compute the band structure including nonlocal exchange effects.
As a result of the reduced Hilbert space of the tight-binding model, in our procedure we neglect the contribution of the double counting of the Fock-like term.
%, needed for the calculation of $L^0$ in Eq.~\eqref{eq:L0_def}, with the perturbative approach detailed in Ref.~\onlinecite{albertoTB}. 
%The sum over states in the e-h propagator $L^0$ and $L^0_{\DFT}$ have been limited to the $\pi$, $\pi^*$ bands and described  with a five nearest-neighbor tight-binding model as described in Ref.~\onlinecite{albertoTB}.
%As nonlocal exchange effects are included only in the $\pi$/$\pi^*$ Hilbert subspace, 
Thus, $\tilde{V}_{\GDFT}=V_{\mathrm{KS}}$ and $\Delta f_{\Hxc}=0$.

We used a $\mathcal{W}$ cutoff in reciprocal space of $4.5$ \AA$^{-1}$. The screened interaction $\mathcal{W}$ is modeled with Dirac-cone screening in the random-phase approximation and calculated self-consistently with the band structure~\cite{albertoTB}.
Doping effects are included both in the occupation factors and the screened interaction $\mathcal{W}$.
The screened matrix elements $V_{\DFT}$ are computed with a first nearest-neighbor tight-binding model with parameters fitted over a DFPT ab-initio calculation, as explained in Ref.~\onlinecite{Venezuela_11}. 

The matrix elements $\Delta C_{s\alpha,r\beta}$ are solved with the Sternheimer equation explained in Ref.~\onlinecite{Giovanni_2024}, with linear mixing and a convergence threshold of the force constants of $10^{-12}$.
We used telescopic k grids~\cite{Binci_2021} that in the dense region around K are equivalent to a $2896\times2896$ uniform grid, while in the coarse region they correspond to a $90\times90$ uniform grid. The parameters are $p=3$, $\mathcal{N}=25$, $l=6$ and $L=10$ in the DFT calculation while $L=11$ in the tight-binding calculations.
The electronic temperature is set to $300$ K in Fig. 1 of the main text and  $70$ K in all the others. The same results at 300 K are presented in the SI.

We calculate the phonon frequency and full-width half maximum (FWHM) from the peak position and width of the spectral function
\begin{equation}
A(\omega)={-}\sum_{s\alpha}{\rm Im}\left[\frac{\omega}{\pi} G_{s\alpha,s\alpha}(\omega)\right] .
\end{equation}
%Phonon frequencies and linewidth are obtained as the peak position and full-width half maximum of the spectral weight (Eq. (36) del paper di Giovanni)
%Mettiamo parte immaginaria
%the real and imaginary parts of the solution of the secular equation $D(\Omega_{\lambda})e_{\lambda} = \Omega_{\lambda}e_{\lambda}$. .
Dynamical effects are computed in the neighborhood of $\Gamma$ and K evaluating the dynamical matrix at the static frequencies $\omega_{\Gamma}$ and $\omega_{\textrm{K}}$.
Due to the finite k-point mesh, we replace the $0^+$ limit in the definition of the phonon Green's function with a finite damping $\eta = 10$ meV.
\bibliography{bibliography}